\documentclass[%
 reprint,
superscriptaddress,
 amsmath,amssymb,
 aps,
 cite,
floatfix,
]{revtex4-1}

\usepackage{graphicx}% Include figure files
\usepackage{dcolumn}% Align table columns on decimal point
\usepackage{bm}% bold math
\usepackage{color}

\begin{document}

\preprint{APS/123-QED}

\title{Prediction of Cross-Fitness for Adaptive Evolution to Different Environmental Conditions: Consequence of Phenotypic Dimensional Reduction}% Force line breaks with \\

\author{Takuya U. Sato\footnotetext\footnotetext}
    \email{takuya.sato.zs@riken.jp}
    \affiliation{Center for Biosystems Dynamics Research (BDR), RIKEN, 6-2-3 Furuedai, Suita, Osaka 565-0874, Japan}
    \affiliation{Universal Biology Institute, School of Science, The University of Tokyo, Faculty of Science Bldg.1, 7-3-1 Hongo, Bunkyo-ku, Tokyo 113-0033, Japan.}
\author{Chikara Furusawa\footnotetext\footnotetext}
    \email{chikara.furusawa@riken.jp}
    \affiliation{Center for Biosystems Dynamics Research (BDR), RIKEN, 6-2-3 Furuedai, Suita, Osaka 565-0874, Japan}
    \affiliation{Universal Biology Institute, School of Science, The University of Tokyo, Faculty of Science Bldg.1, 7-3-1 Hongo, Bunkyo-ku, Tokyo 113-0033, Japan.}
\author{Kunihiko Kaneko\footnotemark\footnotemark}%
    \affiliation{Center for Complex Systems Biology, Universal Biology Institute, University of Tokyo, Komaba, Tokyo 153-8902, Japan}
    \affiliation{The Niels Bohr Institute, University of Copenhagen, Blegdamsvej 17, Copenhagen, 2100-DK, Denmark}
 \email{lapikaneko@gmail.com}

\date{\today}% It is always \today, today,

\begin{abstract}

    How adaptive evolution to one environmental stress improves or suppresses adaptation to another is an important problem in evolutionary biology. 
    For instance, in microbiology, the change of resistance to one antibiotic by resistance acquisition by another drug is a critical issue that has been investigated as cross-resistance. 
    Recent experiments on bacteria have suggested that the cross-resistance of their evolution to various stressful environments can be predicted based on the transcriptome changes after evolution under the corresponding stresses.
    However, there are no studies so far that explain a possible theoretical relationship between cross-resistance and changes in the transcriptome, which causes high-dimensional changes to cell phenotype.
    Here, we show that a correlation exists between fitness change in stress tolerance evolution and response to the environment, using a cellular model with a high-dimensional phenotype and establishing the relationship theoretically.

    In the present study, we numerically evolved a high-dimensional gene regulation network, where genes determined the network and gene expression dynamics. 
    From this network, the fitness of cells under given environmental conditions was determined by expression patterns of output genes.
    By numerically evolving such cells to satisfy several relationships between environmental inputs and protein expression outputs, we demonstrated that evolutionary changes in phenotypes are constrained to a low-dimensional subspace, whose dimension is given by the number of input-output relationships.
    This dimensional reduction is explained by the separation of a few eigenvalues in the Jacobian matrix for the expression dynamics. Additionally, we applied a large variety of environmental changes to these evolved cells and examined further evolution under these stress conditions. 
    The correlation of the evolutionary change in fitness in response to different stresses is well predicted by the non-evolutionary environmental responses to each stress. 
    This prediction is indeed possible, as evolutionary changes and adaptive responses are within the same constrained subspace. 
    Therefore, by taking advantage of dimensional reduction, we formulated a potential theory of phenotypic changes caused by environmental and genetic changes. 
    Then, the cross-fitness to different environmental stresses was computed by the Hessian matrix of the potential, which supports the results of the numerical evolution. 
    Finally, we applied the theory to experimental data on bacterial evolution under antibiotics.
    The data demonstrate the predicted correlation between the fitness changes by evolution and transcriptome changes upon environmental stresses.
    The present results allow for the prediction of evolution from transcriptome information in response to different stresses before evolution.
    The relevance of this to microbiological evolution experiments is discussed.
    
    % \begin{description}
    %     \item[Usage]
    %     Secondary publications and information retrieval purposes.
    %     \item[Structure]
    %     You may use the \texttt{description} environment to structure your abstract;
    %     use the optional argument of the \verb+\item+ command to give the category of each item. 
    % \end{description}
\end{abstract}

\maketitle

%\tableofcontents

\section{Introduction}
\label{sec:Intro}

    Generally, organisms change their state to adapt to various environmental stresses.
    This ability is thought to have been acquired through evolution \cite{zhang2011acceleration, watkins2016overview}.
    Those who evolved to adapt to one environment may increase or decrease the degree of their adaptation to another environment.
    For example, adaptive evolution to one stressful environment may increase or decrease fitness to manage another stress as compared with that of the organism before evolution.
    This correlated change in fitness is called 
    cross-resistance \cite{gilbert2003potential,levy2004antibacterial,gnanadhas2013biocides,lazar2013bacterial,imamovic2013use,suzuki2014prediction,sommer2017prediction,suzuki2017acceleration,rodriguez2015collateral}.
    If the adaptive evolution to one environmental stress increases or decreases the fitness for another, the cross-resistance is positive or negative, respectively.
    In medicine, understanding the cross-resistance of bacteria to different antibiotics is a crucial issue.
    
    Can such cross-resistance be predicted?
    Extensive studies have been conducted to uncover specific genetic mutations which allow adaptive evolution to individual environmental stresses and to unveil functional changes that occur as a result of such mutations.
    Molecular changes caused by mutations have been identified in certain genes, which allow for resistance to environmental stress\cite{toprak2012evolutionary,lopatkin2021clinically}.
    Molecular changes caused by mutations have been identified in certain genes, which allow for resistance to environmental changes.
    However, the detailed mechanisms of cross-resistance remain unclear.
    Cross-resistance between different environmental conditions involves interactions among diverse components that are influenced by the mutation and are not explained directly by specific molecular changes.
    Examination of the correlation between fitness changes across different environmental conditions using standard molecular biology methods that focus on a one-to-one correspondence between genes and functions is not easy.
    
    How can we compare adaptive evolution under different environmental conditions?
    For this purpose, we need to consider changes to the cellular state that is shaped by a wide variety of components. 
    Such a cellular state can be represented by the concentrations of these components.
    Changes in the cellular state in response to environmental changes, such as antibiotics, temperature, and nutritional conditions, will lead to a change in the growth of a cell.
    The correlation of changes in the cellular state across different environmental changes will provide information on how organisms evolve to them.
    Such information involves high-dimensional data that characterize the cellular state.
    
    Recent advances in experimental techniques have enabled the acquisition of high-dimensional data of cellular states, such as the transcriptome, proteome, and metabolome \cite{taniguchi2010quantifying,han2006escherichia,yuan2009metabolomics}.
    Using these high-dimensional data, a detailed analysis of the cellular state is now possible.
    However, how can we extract relevant information from high-dimensional data with thousands of components to obtain the correlation between evolutionary adaptation to different conditions?
    
    A recent experimental report examined transcriptome changes throughout the evolution of bacteria in response to a variety of environmental stresses \cite{suzuki2014prediction,horinouchi2017prediction}.
    In these studies, the authors measured the cross-resistance, that is, how adaptive evolution to one environment, $E_1$, changed the growth rate of bacteria in another environment, $E_2$.
    Then, by measuring transcriptome changes through adaptive evolution, they constructed a low-dimensional linear model for these changes, explaining the observed cross-resistance.
    Notably, the environmental stresses adopted in their experiments had a variety of molecular effects on cells.
    The transcriptome of \textit{E. coli} used in their experiment was high-dimensional data with over 4000 dimensions.
    Despite this complexity, low-dimensional information extracted from high-dimensional information is suggested to be relevant to predict cross-resistance to a variety of conditions to a certain degree.
    
    If cellular states moved throughout the entire high-dimensional space during adaptive evolution to various stress environments, changes in phenotype (i.e., cellular state) in response to different stress environments would not be correlated, and predictions of cross-resistance by the environmental response would not be possible.
    However, such predictions may be possible if transcriptome changes due to adaptive evolution are restricted to a relatively low-dimensional subspace.
    Is there general support for a such low-dimensional reduction in adaptive changes to cellular states?
    
    Several recent experiments have suggested that changes in cellular state in response to environmental stresses are constrained in low-dimensional space \cite{horinouchi2010transcriptome,carroll2013evolution,keren2013promoters,kaneko2015universal,horinouchi2015phenotypic,stolovicki2011collective}.
    Changes in the transcriptome of \textit{E. coli} across various stress environments were found to be strongly correlated.
    Horinouchi et al. also showed that transcriptomic changes in independent evolutionary lineages converge along the common principal component (PC) space in the adaptive evolution of \textit{E. coli} under ethanol stress.
    These results suggest that phenotypic changes in the adaptation and evolution of cells in response to environmental stress occur within a low-dimensional space.
    
    How are phenotypic changes constrained to a low-dimensional space?
    By simulating a catalytic chemical reaction network model with thousands of components, it was found that high-dimensional concentration changes in response to environmental or mutational changes are constrained to a common low-dimensional space as a result of evolution to increase the fitness \cite{furusawa2015global,furusawa2018formation,sato2020evolutionary}.
    This constraint is then formulated in terms of dynamical systems theory as a separation of a few slow eigenmodes for the relaxation dynamics of the rate equation representing the cellular state changes.
    
    Can we, then, theoretically predict cross-resistance using the information in such low-dimensional constraints \cite{tikhonov2020model}?
    In the presence of phenotypic constraints, responses to environmental and evolutionary changes are restricted to a common, lower-dimensional subspace.
    Accordingly, one does not need the entire high-dimensional data to predict the fitness change; information within the low-dimensional subspace will be sufficient to estimate the fitness changes across environmental conditions.
    Thus, the information needed to predict cross-resistance is significantly reduced.
    In the present study, we used a gene regulatory network (GRN) model to demonstrate such low-dimensional phenotypic constraints by evolution, and then demonstrated that cross-resistance is predicted by cellular responses to stress before evolution by taking advantage of phenotypic constraints \cite{glass1973logical,mjolsness1991connectionist,salazar2001phenotypic,kaneko2007evolution,inoue2021entangled,nagata2020emergence}.
    
    The remainder of this paper is organized as follows:
    In Sec.\ref{sec:model}, we introduce the GRN model of a cell used in the present study and describe the procedure of simulated evolution.
    Next, in Sec.\ref{sec:phenotypic_constraint_grn}, we show that phenotypic constraints are produced when the GRNs are evolved under fitness to satisfy multiple input-output relationships.
    We demonstrate that the degree of the phenotypic constraint acquired through evolution is determined by the number and strength of the postulated input-output relationships.
    We also explain such constraints in terms of the nature of gene regulatory matrices.
    In Sec.\ref{sec:prediction_cross_resistance}, we show the results of simulations of adaptive evolution to a variety of environmental stresses, by using the evolved GRNs obtained in Sec.\ref{sec:phenotypic_constraint_grn} as the ancestor.
    Then, we computed the cross-fitness, that is, the fitness of a cell that has evolved under another environment for a new environment.
    We demonstrated that this cross-fitness is approximated using low-dimensional variables along with phenotypic constraint coordinates.
    In particular, when a $P$-dimensional phenotypic constraint exists, the cross-fitness and cross-resistance that are desired from it are approximately described by a function of $P$ variables.
    Then, the cross-fitness as a result of evolution is predicted by the correlation in transcriptome changes upon environmental stresses.
    In Sec.\ref{sec:potential_approximation_to_cross_fitness}, the approximate form of cross-fitness in Sec.\ref{sec:prediction_cross_resistance} is derived by assuming that fitness is given by a potential function of low-dimensional environmental and genetic coordinates. 
    In Sec.\ref{sec:application_for_experimental_data}, we apply the present theory to experimental data of evolution of resistance to antibiotics in \textit{E.coli}.
    The experimental data well reproduce the predicted correlation between the cross-fitness and transcriptome changes.
    In Sec.\ref{sec:discussion}, we summarize the result and discuss its relevance to cross-resistance observed in experiments of bacterial evolution of antibiotic resistance.

\section{Model}
\label{sec:model}

    \subsection{Cell Model}
    \label{ssec:cell_model}
    
        We adopted a GRN as a model for the cellular state.
        The GRN is composed of $N$ genes whose expression is represented by the $N$-dimensional vector $\boldsymbol{x}=(x_1,x_2,\cdots,x_N)$, and the cell state is given by this vector.
        The time evolution of the state follows the rate equation:
        \begin{subequations}
            \begin{align}
                \dot{x_i} &= f(y_i) - x_i, \label{eq:rate_equation}\\
                f(y_i) &= \frac{1}{1+\exp(-y_i)},\\
                y_i &= \frac{1}{\sqrt{N}}\sum_{j=1}^N G_{ij}x_j + \frac{1}{N_I}\sum_{j=1}^{N_I}I_{ij}\eta_j+E_i.
            \end{align}
        \end{subequations}
        $\boldsymbol{G}$ is an $N\times N$ matrix representing the interactions between genes, satisfying $G_{ij}\in\{-1, 1\}\ (i\neq j),\ G_{ij}=\ (i=j)$.
        $G_{ij} > 0 $ indicates that the product of $j$th gene positively regulates the $i$th gene, that is, accelerates its transcription. $G_{ij}<0$ represents negative regulation.
    
        $\boldsymbol{\eta}$ is an $N_I$-dimensional vector that represents the input signal from the external environment to the cell, satisfying $\eta_i \in \{-1, 1\}$.
        The strength of the interactions between the input signal and the GRN is represented by the $N \times N_I$ matrix $\boldsymbol{I}$, satisfying $I_{ij}\in\{-1, 1\}$.
        $\boldsymbol{E}$ is an $N$-dimensional vector representing the environmental stress.
        In the parametric region used in this study, cellular states always reach a unique fixed point $\boldsymbol{x^*}=(x^*_1, x_2^*,\dots,x_N^* )$ as a result of time evolution using the rate equation (Eq.(\ref{eq:rate_equation})).
        In this study, we refer to the fixed point $\boldsymbol{x^*}$ of Eq.(\ref{eq:rate_equation}) as the phenotype. 
        The phenotype $\boldsymbol{x^*}$ is uniquely determined for genotype $\boldsymbol{(I, G, O)}$ and environment $\boldsymbol{\eta}, \boldsymbol{E}$.
    
        Both terms $\boldsymbol{I\eta}$ and $\boldsymbol{E}$ represent the interactions between the external environment and the GRN, but their biological meanings are different.
        $\boldsymbol{I\eta}$ represents the signal inputs from the external environment.
        Such input from the environment appears frequently over long-term, evolutionary timescales, allowing cells to adapt to these environments through evolution. 
        For such evolved cells, we applied environmental stress $\boldsymbol{E}$ for a laboratory timescale, much smaller than the long-term evolutionary timescale (consider, for instance, the application of antibiotics to wild-type bacteria).
        Against such inputs, cells may be required to evolve by transient adaptations, which are lost in the long-term evolutionary time scale.
        
        In this model, the fitness of a cell is determined by the expression of the output genes, that is, the vector $\boldsymbol{o}=(o_1,o_2,\dots,o_{N_O})$.
        The stationary expression of the output genes is given by $o_i^*=f(\sum_{j=1}^NO_{ij}x_j^*/\sqrt{N})$, where $\boldsymbol{O}$ is an $N_O \times N$ matrix of interactions between the genes in GRN and the target gene, satisfying $O_{ij}\in\{-1, 1\}$.
        Here, we postulate that the fitness for each condition $\boldsymbol{\eta^{(n)}}$ is defined by the negative distance $-|\boldsymbol{o^*}-\boldsymbol{t^{(n)}}|^2$ between the output gene expression and the optimal gene pattern $\boldsymbol{t^{(n)}}$ corresponding to each input signal $\boldsymbol{\eta^{(n)}}$, that is, the fitness takes a maximum value of zero if the expression pattern of the output genes $\boldsymbol{o}$ matches the optimal gene pattern $\boldsymbol{t^{(n)}}\ (n=0,1,\dots, P-1)$.
        Now, we consider $P$ different environmental conditions with input signal $\boldsymbol{\eta^{(n)}} \ (n=0,1,2,\dots,P-1)$ and an optimal gene pattern $\boldsymbol{t^{(n)}}$.
        % Then, the fitness is given by $-\sum_m(\boldsymbol{o^*}|_{\eta=\eta^{(n)}}-\boldsymbol{t^{(n)}})^2$.
        In this study, we consistently use $N = 100$, $N_I = 8$, and $N_O = 8$.

    \subsection{Evolution}
    \label{ssec:method_evolution}
        Evolutionary simulations were performed using the following procedure.
        In each generation, $M$ mutant cells were created from $L$ mother cells.
        The total population was $ML$.
        Mutant cells were generated by reversing the sign of each matrix element of the genotype $(\boldsymbol{I, G, O})$ of the mother cell with probability $\rho$.
        In this study, $\rho = 1/N^2 = 0.0001$ was used.
        The fitness of each mutant cell was then calculated as follows:
        We calculated the fixed points $\boldsymbol{x^*}$ and $\boldsymbol{o^*}$ using the rate equation Eq.(\ref{eq:rate_equation}) using the 4-degree adaptive Runge-Kutta method \cite{press1992adaptive}, and used this to calculate the fitness.
        Initial states for the calculation of the fixed point were randomly chosen from the uniform distribution $0<x_i<1\ (i=1,2,\dots, N)$.
        However, because the model adopted in the present study has only one fixed point in the parameter region, the choice of initial values does not affect the results.
        Finally, the top $L$ fitted cells were selected for the next generation of mother cells. 
        In this study, we use $M=4$ and $L=25$.

\section{Evolutionary dimension reduction in the Gene Regulatory Network}
\label{sec:phenotypic_constraint_grn}
    
    \subsection{Fitness}
    \label{ssec:fitness_dr_evolution}

        First, we performed evolution from randomly generated cells with given matrices $\boldsymbol{I^{ini}}, \boldsymbol{G^{ini}}$ and $\boldsymbol{O^{ini}}$.
        $\boldsymbol{I^{ini}}, \boldsymbol{G^{ini}}$ and $\boldsymbol{O^{ini}}$ are randomly generated with probability $p$ to take $\pm 1$ as follows:
        \begin{subequations}
            \begin{align}
                p(I^{ini}_{ij}=\pm 1) &= \frac{1}{2},\\
                p(G^{ini}_{ij}=\pm 1) &= \frac{1}{2},\\
                p(T^{ini}_{ij}=\pm 1) &= \frac{1}{2},
            \end{align}
        \end{subequations}
        whereas $G_{ii}$ is set to 0.
    
        As mentioned, we assumed that cells need to respond appropriately to external inputs to survive; as such, output genes should take the appropriate expression pattern $\boldsymbol{t^{(n)}}$ upon input signal $\boldsymbol{\eta^{(n)}}$.
        Fitness for a input-output pair $(\boldsymbol{\eta^{(n)}}, \boldsymbol{t^{(n)}})$ under environmental stress $\boldsymbol{E}$ is given as followed;
        
        \begin{equation}
            \mu_n(\boldsymbol{E}) = -\sum_{i=1}^{N_O}\left|o_i^*|_{\boldsymbol{\eta^{(n)}},\boldsymbol{E}}-t_i^{*(n)}\right|,
            \label{eq:mu_m}
        \end{equation}
        where $\boldsymbol{o}^*|_{\boldsymbol{\eta^{(n)}}}$ is stationary expression pattern of output genes with input signal $\boldsymbol{\eta^{(n)}}$ and environmental stress $\boldsymbol{E}$.
        Note that $\mu_n(\boldsymbol{E}) \leq 0$ and $\mu_m(\boldsymbol{E})$ takes $0$ only if the stationary expression pattern of output genes agrees with the target pattern.
        
        In this section, by considering $P$ input-output relationships without environmental stress, that is $\boldsymbol{E=0}$, we used the following fitness function $\bar{\mu}$:
    
        \begin{equation}
            \bar{\mu}=\frac{1}{P}\sum_{n=1}^P\mu_n(\boldsymbol{E=0}).
        \end{equation}
    
        $\bar{\mu}$ takes a maximum value of 0 only when the output gene expression pattern agrees with the target pattern $\boldsymbol{t^{(n)}}$ for each of the input signals $\boldsymbol{\eta^{(n)}}\ (n=0,1,\dots, P-1)$ from the environment.
        In this study, we used a non-signal condition and the following $\tilde{P}$ pairs of  signals and expression patterns of the output genes (i.e. $P=2\tilde{P}+1$):
        \begin{subequations}
            \begin{align}
                \eta^{(0)}_i=0\ &,\ t^{(0)}_i=1/2,     \label{eq:inputoutput1}\\
                \eta^{(2\tilde{P}-1)}_i\in\{-1,1\}\ &,\ t^{(2\tilde{P}-1)}_i\in\{\frac{1-2\alpha}{2},\frac{1+2\alpha}{2}\},\label{eq:inputoutput2}\\
                \eta^{(2\tilde{P})}_i=-\eta^{(2\tilde{P})}_i\ &,\ t^{(2\tilde{P})}_i=1-t^{(2\tilde{P})}_i.\label{eq:inputoutput3}
            \end{align}
        \end{subequations}
        
\begin{figure}[hbt]
            \centering
            \includegraphics[width=\hsize]{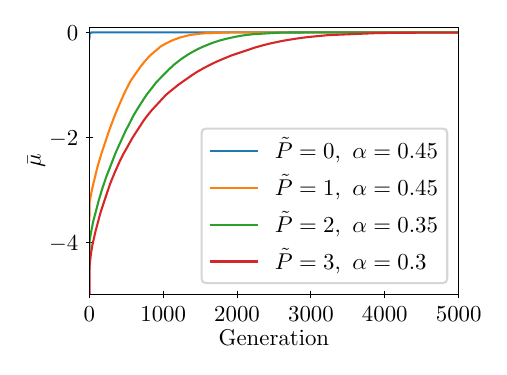}
            \caption{The evolutionary increase of the given fitness $\bar{\mu}$, averaged over population under signal-target relationships.
            The cases $(\tilde{P}=0, \alpha=0.45), (\tilde{P}=1, \alpha=0.45), (\tilde{P}=2, \alpha=0.35)$ and $(\tilde{P}=3, \alpha=0.3)$ are plotted.}
            \label{fig:fitness}
        \end{figure}

        Here, $\alpha$ is a parameter that represents the strength of the required output gene response and we define $\boldsymbol{\alpha^{(n)}}$ which satisfies $t^{(n)}_i=(1+2\alpha^{(n)}_i)/2$.
        The first signal-target relationship $(\boldsymbol{\eta^{(0)}}, \boldsymbol{t^{(0)}})$ requires that there is no response to output genes when there are no environmental signals.
        For $\tilde{P}\geq1$, we set a pair of patterns $2\tilde{P}-1$ and $2\tilde{P}$ is symmetric from the case with no input signal.
        In addition, each input pattern $\boldsymbol{\eta^{(2n)}}\ (n=1,2, \dots, \tilde{P})$ is chosen to be linearly independent.
        The $\boldsymbol{\alpha^{(2n)}}\ (n=1,2,\dots,\tilde{P})$ is linearly independent.
        That is, $(\boldsymbol{\eta^{(2m)}}\cdot\boldsymbol{\eta^{(2n)}}) = 0\ (m \neq n)$ and $(\boldsymbol{\alpha^{(2m)}} \cdot \boldsymbol{\alpha^{(2n)}})=0\ (m \neq n)$.
        When a set of $(\boldsymbol{\eta^{(n)}}, \boldsymbol{t^{(n)}})\ (n=0,1,\dots,2\tilde{P})$ is given by the above methods, there are $\tilde{P}$ signal-target relationships.
        The purpose of the above pairwise signal-target relationship is to ensure that the symmetry of the phenotypic constraints is obtained as a result of evolution.
        However, the phenotypic constraints discussed below are obtained even when the signal-target relationship is randomly assigned, without the above symmetry.
        As a result of evolution, the fitness approached maximum $\bar{\mu} \sim 0$, with $\mu_n(\boldsymbol{E}=0) \sim 0$ for $n\leq2\tilde{P}$, as long as $\tilde{P}$ and $\alpha$ are not so large(Fig.\ref{fig:fitness}).
        In the following sections, we study the behavior of such evolved networks.
               
    \subsection{Evolutionary dimension reduction}
    \label{ssec:phenotypic_change}
        
\begin{figure*}[hbt]
            \centering
            \includegraphics[width=\hsize]{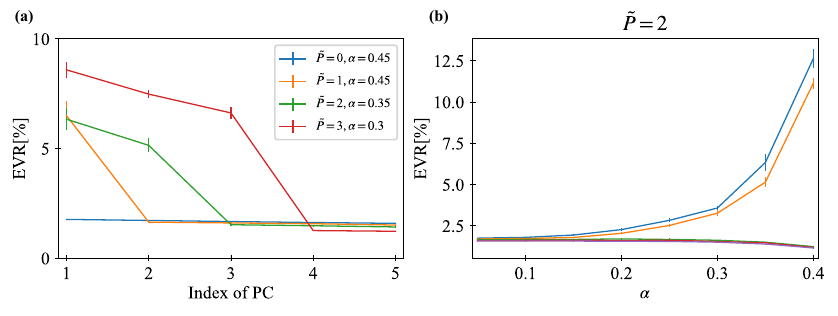}
            \caption{Explained variance ratio (EVR) of phenotypic changes for the first 5 principal components (PCs) when random environmental stresses $\boldsymbol{E}$ generated by $E_i\sim N(0,1)$ were applied.
            The changes in $\boldsymbol{x^*}$ for the evolved gene regulatory network were computed.
            The phenotypic changes $\boldsymbol{\delta x^*(E)}$ were obtained for 10,000 independent environmental stresses, from which PCs were computed.
            (a) The explained variance ratios (EVR) of the PCs of the phenotypic changes of the cell which evolved under different $\tilde{P}$ input-target relationships are plotted.
            Each explained variance ratio was calculated for the cell with the top fitness value in the population that evolved for $(\tilde{P},\alpha) = (0,0.45),(1,0.45),(2,0.35),(3,0.3)$.
            (b) The explained variance ratios are plotted against the strength of the required target gene response $\alpha$.
            Each explained variance ratio was calculated for the cell with the top fitness value in the population that evolved for $\tilde{P}=1$.
            Please also see supplementary figure Fig.S1 about the cases $\tilde{P}=0,1,2,3$.
            The error bars represent the standard deviation of the five independent strains.}
            \label{fig:Var_candidate_env}
        \end{figure*}

        The phenotype of cell $\boldsymbol{x^{*}}$ changes when environmental stress $\boldsymbol{E}$ is imposed.
        We denote the phenotypic change in response to environmental stress as $\boldsymbol{\delta x^{*}(E) = x^{*}(E) - x^{*}(0)}$, where $\boldsymbol{x^*(E)}$ represents the phenotype of the cell under environmental stress $\boldsymbol{E}$, calculated using the environmental signal $\boldsymbol{\eta^{(0)}}$.
        We calculated phenotypic changes $\boldsymbol{\delta x^{*}(E)}$ for cells evolved under various $\tilde{P}$ and $\alpha$, subjected to 10,000 randomly generated environmental stresses $\boldsymbol{E}$.
        These environmental stresses $\boldsymbol{E}$ were generated such that each element followed a normal distribution with a mean of 0 and a variance of 1. 
        We investigated the change in phenotype with environmental stress $\boldsymbol{\delta x^*(E)}$ in the $N$-dimensional phenotypic space.
        However, as it is too high-dimensional, we performed principal component analysis (PCA) of over 10,000 phenotypic changes $\boldsymbol{\delta x^*(E)}$ and examined if the variance was explained by a few components.
        The dependence of the explained variance on $\tilde{P}$ and $\alpha$ is illustrated in Fig.\ref{fig:Var_candidate_env}.

        To study the validity of dimension reduction, we examined the dependence of the explained variance ratio (EVR) on $\tilde{P}$ and $\alpha$.
        As shown in Fig.\ref{fig:Var_candidate_env}(a), the contribution of the top $\tilde{P}$ PCs are large, whereas the components beyond $\tilde{P}$ remain small.
        Recall that $\alpha$ is a parameter that represents the strength of the required target gene response; the larger $\alpha$, the larger the response.
        The top $\tilde{P}$ PCs account for a larger portion of the phenotypic change $\boldsymbol{\delta x^{*}(E)}$ than other PCs do.
        This result implies that $\tilde{P}$, which represents the number of independent signal-target relationships, determines the dimension of the phenotypic constraint.
        In contrast to the one-dimensional constraint studied earlier \cite{kaneko2015universal, furusawa2018formation, sakata2020dimensional}, the constraint to $\tilde{P}(>1)$-dimensional constraint is generated, corresponding to the degree of freedom of environmental conditions in which the adaptive evolution progressed \footnote{In previous studies \cite{furusawa2018formation,sakata2020dimensional} using catalytic chemical reaction networks, one-dimensional phenotypic constraints correlated with growth rate were acquired even in evolution among multiple environments. 
        In the present paper, a gene regulatory network model that does not include growth rate was used to consider higher dimensional phenotypic constraints.}.

        In summary, the dimension of the phenotypic constraint agrees with the degrees of freedom $\tilde{P}$ of the signal-target relationship, and the magnitude of the variance in these directions is correlated with the magnitude of the required target response. 
        Note that the environmental stresses adopted to compute phenotypic variations are not included in the environment where evolution has taken place. 
        However, the response to novel environmental changes is restricted to $\tilde{P}$-dimensional space after evolution.
        We also observed this in phenotypic changes caused by genetic mutations (Fig.S2) in the high-dimensional gene expression space and the corresponding dynamical system analysis for an origin of phenotypic constraint in the dynamical system (see Sec.S3, S4, and S5 in supplementary material).
        These phenotypic changes due to environmental stresses and genetic mutation are restricted to a common low-dimensional space.
        This will be important for the correspondence  between  phenotypic changes in response to environmental stresses and due to adaptive evolution, to be studied in the following sections.

\section{Prediction of cross-resistance by phenotypic constraint}
\label{sec:prediction_cross_resistance}

    \subsection{Fitness}
    \label{ssec:fitness_under_stress}

        In the previous section, we numerically evolved cells to realize the appropriate target pattern $\boldsymbol{t^{(n)}}\ (n=0,1,\dots,2\tilde{P})$ in response to each input signal $\boldsymbol{\eta^{(n)}}\ (n=0,1,\dots,2\tilde{P})$.
        As a result, phenotypic changes $\boldsymbol{\delta x^*}$ in response to environmental stress $\boldsymbol{E}$ and mutation to genotype $\boldsymbol{G}$ were constrained to the same subspace with $\tilde{P}$ dimensions.

        In this section, we adopt cells that have already evolved as in the previous section, achieved the phenotypic constraint, and then studied the evolution of adaptation to novel environmental stresses.
        This corresponds to the short-term adaptive evolution in laboratory experiments.
        Using this setup, we computed the cross-fitness, that is, the fitness of a cell that has evolved to adapt to an environmental stress $\boldsymbol{E^{(1)}}$, exposed to another environmental stress $\boldsymbol{E^{(2)}}$, and we show that the cross-fitness can be predicted by phenotypic constraints in a low-dimensional subspace.

        In this section, we used the fitness $\mu_0(\boldsymbol{E})$ (see Eq.\ref{eq:mu_m}) in the evolutionary simulations.
        Evolution with this fitness requires that the appropriate target pattern $\boldsymbol{t^{(0)}}$ be realized in response to the input signal $\boldsymbol{\eta^{(0)}}$ in the presence of environmental stress $\boldsymbol{E}$.
        Although $\boldsymbol{\eta^{(0)}}$ and $\boldsymbol{t^{(0)}}$ were used here, the qualitative results did not change when the other pairs of input signals and target gene response patterns were adopted.
        We calculated the fitness with one input-target relationship, assuming evolution under a constant environment over a short period, such as laboratory evolution.

    \subsection{cross-fitness}
    \label{ssec:cross_fitness}

        \begin{figure}[hbt]
            \centering
            \includegraphics[width=\hsize]{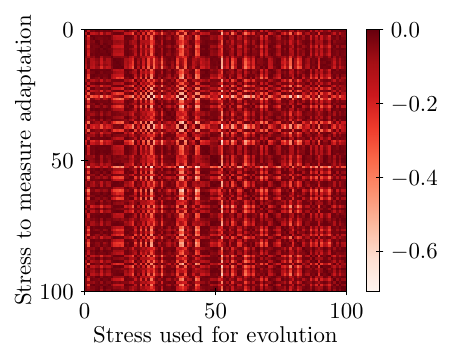}
            \caption{Fitness when cells evolved under given stress types and are exposed to different stress types. 100 environmental stress types $(\boldsymbol{E^{(1)}}, \boldsymbol{E^{(2)}},\dots, \boldsymbol{E^{(100)}})$ were randomly generated. The vertical axis represents the environmental stress used to measure adaptation, and the horizontal axis represents the environmental stress used for evolution.
            Stresses are ordered by the number of random seeds used to generate environmental stress.
            Here, $y_1(\boldsymbol{E^{(i)}})$ is the phenotypic change of $\boldsymbol{\delta x^*(E^{(i)})}$ when environmental stress $\boldsymbol{E^{(i)}}$ is applied to the cells before evolution.}
            \label{fig:heatmap}
        \end{figure}

        Now, we introduce the cross-fitness $\mu_{cross}(\boldsymbol{E^{(1)}}, \boldsymbol{E^{(2)}})$ which is defined as the fitness of genotype $\boldsymbol{G^*(E^{(1)})}$, that is, the fitness when the cell, which has evolved to adapt to environmental stress $\boldsymbol{E^{(1)}}$, is exposed to environmental stress $\boldsymbol{E^{(2)}}$.
        Thus, it is represented as 
    
        \begin{equation}
            \mu_{cross}(\boldsymbol{E^{(1)}},\boldsymbol{E^{(2)}}) = \mu_0(\boldsymbol{E^{(2)}})_{\boldsymbol{G=G^*(E^{(1)})})} .
        \end{equation}
    
        In other words, the cross-fitness $\mu_{cross}(\boldsymbol{E^{(1)}},\boldsymbol{E^{(2)}})$ represents the degree of adaptation under environmental stress $\boldsymbol{E^{(2)}}$ of the cells that evolved to adapt to a different environmental stress $\boldsymbol{E^{(1)}}$.
    
        In Fig.\ref{fig:heatmap}, $\mu_{cross}(\boldsymbol{E^{(1)}},\boldsymbol{E^{(2)}})$ is plotted as a heat map with the evolved environment $\boldsymbol{E^{(2)}}$ as the horizontal axis and the environment $\boldsymbol{E^{(1)}}$ used to measure the fitness as the vertical axis.
        From the figure, it is difficult to obtain information from the heat map in which the environmental stresses $\boldsymbol{E}$ are randomly ordered.
    
        To predict cross-fitness, we must find an appropriate feature variable $\boldsymbol{y(E)}$ that captures the effective internal state corresponding to environmental stress $\boldsymbol{E}$.
        $\boldsymbol{y(E)}$ is a quantity determined by the cellular state before evolution to adapt to the stress.
    
\begin{figure*}[hbt]
            \centering
            \includegraphics[width=\hsize]{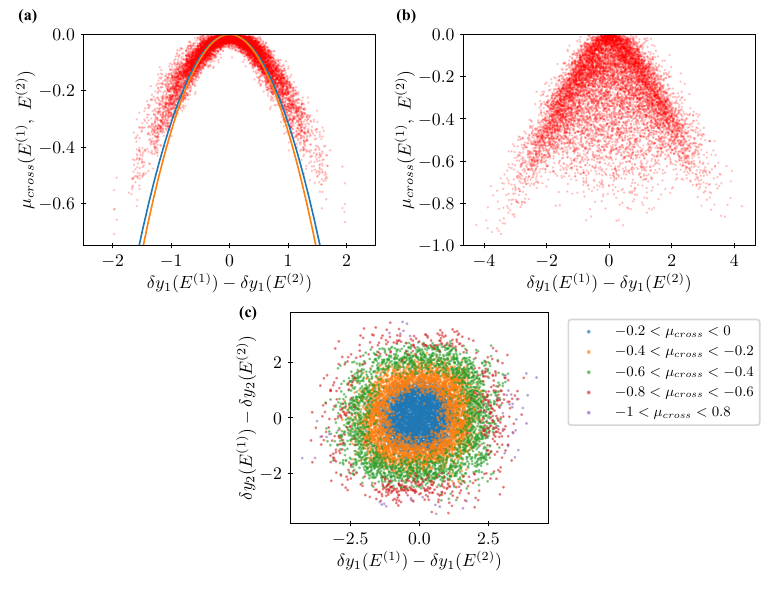}
            \caption{Cross-fitness $\mu_{cross}(\boldsymbol{E^{(1)}},\boldsymbol{E^{(2)}})$ is plotted against the difference between the feature values $\boldsymbol{y(E^{(1)})-y(E^{(2)})}$ of environmental stress.
            The feature value $\boldsymbol{y(E)}$ is the principal component vector of principal component analysis.
            (a)$\tilde{P}=1$. The solid lines represent the second-order approximation curve predicted from the theory.
            The second-order coefficient of the blue line is calculated by the least square method from the data on phenotypic changes under 10,000 randomly chosen environmental stresses.
            The second-order coefficient of the orange line is the $(\partial^2 \mu_0/\partial y_1^2)$, calculated using the information for the pre-evolutionary genotypes as given in Sec.\ref{ssec:Application_of_potential_approximation_to_the_result_of_evolutional_simulation}.
            (b)$\tilde{P}=2$. 
            When two-dimensional phenotypic constraints exist, we cannot approximate the cross-fitness with the function of a one-dimensional feature value.
            (c)$\tilde{P}=2$. The pair of environments $(\boldsymbol{E^{(1)}, E^{(2)}})$ used to measure adaptive evolution and cross-adaptation degree is transformed into a two-dimensional feature value space $(y_1(\boldsymbol{E^{(1)}})-y_1(\boldsymbol{E^{(2)}}), y_2(\boldsymbol{E^{(1)}})-y_2(\boldsymbol{E^{(2)}}))$, plotted with colors coded according to the cross-adaptation $\mu_{cross}(\boldsymbol{E^{(1)}, E^{(2)}})$.
            It can be seen that the pairs of environments corresponding to different adaptations are distributed in a doughnut shape.
            This is because the cross-fitness $\mu_{cross}(\boldsymbol{E^{(1)}, E^{(2)}})$ at $\tilde{P}=2$ can be approximated by a monotone univalent function whose arguments are $(y_1(\boldsymbol{E^{(1)}})-y_1(\boldsymbol{E^{(2)}}), y_2(\boldsymbol{E^{(1)}})-y_2(\boldsymbol{E^{(2)}})$.}
            \label{fig:Kernel}
        \end{figure*}

        Here, as a possible candidate for $\boldsymbol{y(E)}$, we adopted the PCs of the phenotypic changes $\boldsymbol{\delta x^*(E)}$ against environmental stress $\boldsymbol{E}$ for the cells before the evolution because the dominant $\tilde{P}\ $PCs capture the phenotypic change under the phenotypic constraint, to which $\boldsymbol{\delta x^*(E)}$ under environmental stress is restricted.
        The PC space was calculated using 10,000 random environmental stresses $\boldsymbol{E}$, whose elements followed a normal distribution with a mean of 0 and variance of 1.
        The value $y_i(\boldsymbol{E})$ is the $i$\ th principal component value for the phenotypic change $\boldsymbol{\delta x^*(E)}$.
        In Fig.\ref{fig:Kernel}(a), the cross-fitness $\mu_{cross}(\boldsymbol{E^{(1)}},\boldsymbol{E^{(2)}})$ across 10,000 random environments is plotted as a function of $y_1(\boldsymbol{E^{(1)}}) - y_1(\boldsymbol{E^{(2)}})$ by red dots.
        It can be seen that the cross-fitness $\mu_{cross}(\boldsymbol{E^{(1)}},\boldsymbol{E^{(2)}})$ can be approximated by a single curve; that is, the cross-fitness $\mu_{cross}(\boldsymbol{E^{(1)}},\boldsymbol{E^{(2)}})$ is approximately represented by a single function $\tilde{\mu}_{cross}(\delta y_1(\boldsymbol{E^{(1)}}, \boldsymbol{E^{(2)}}))$ with \\
        $\delta y_1(\boldsymbol{E^{(1)}}, \boldsymbol{E^{(2)}}) = y_1(\boldsymbol{E^{(1)}}) - y_1(\boldsymbol{E^{(2)}})$,
        This is possible because of the existence of a 1-dimensional phenotypic constraint, as we adopted $\tilde{P}=1$ in this case.
    
        Then, how can cross-fitness be represented for $\tilde{P}=2$, where the constraint is 2-dimensional?
        Here, we show the results for ancestor cells that evolved with $\tilde{P}=2$ and $\alpha=0.4$.
        In Fig.\ref{fig:Kernel}(b), we plotted the cross-fitness $\mu_{cross}(\boldsymbol{E^{(1)}},\boldsymbol{E^{(2)}})$ against $\delta y_1(\boldsymbol{E^{(1)}},\boldsymbol{E^{(2)}})$, similar to Fig.\ref{fig:Kernel}(a).
        In this case, cross-fitness $\mu_{cross}(\boldsymbol{E^{(1)}},\boldsymbol{E^{(2)}})$ cannot be approximated by a function with a single argument $\delta y_1(\boldsymbol{E^{(1)}}, \boldsymbol{E^{(2)}})$.
        Because the dimension of the phenotype constraint has been increased from 1 to 2, the cross-fitness $\mu_{cross}(\boldsymbol{E^{(1)}},\boldsymbol{E^{(2)}})$ is estimated as a function of a 2-dimensional PC plane in Fig.\ref{fig:Kernel}(c).
        The difference in the colors of the dots in the Figure corresponds to the cross-fitness values.
        It can be observed that the points with the same cross-fitness are distributed in a doughnut shape in the 2-dimensional PC plane.
        Cross-fitness $\mu_{cross}(\boldsymbol{E^{(1)}},\boldsymbol{E^{(2)}})$ is represented by a function of two-dimensional arguments $(\delta y_1(\boldsymbol{E^{(1)}}, \boldsymbol{E^{(2)}}), \delta y_2(\boldsymbol{E^{(1)}}, \boldsymbol{E^{(2)}}))$.
        It is suggested that when a $D$-dimensional phenotypic constraint exists, the cross-fitness $\mu_{cross}(\boldsymbol{E^{(1)}},\boldsymbol{E^{(2)}})$ can be approximated as a function $\tilde{\mu}_{cross}(\delta y_1,\dots, \delta y_D)$.
    
        In this section, we demonstrate the existence of an approximation function for the cross-fitness.
        These results suggest that the response of cells to environmental changes and evolution can be linked by phenotypic constraints, from which we can predict the cross-fitness in terms of a few, that is, $\tilde{P}$, PCs of the phenotypic change before evolution to novel environmental stresses.
        
    \subsection{Prediction of cross-resistance by cosine-similarity}
    \label{ssec:prediction_cross-resistance}

        So far, we have shown that cross-fitness can be approximated by a single curved surface using the information on the phenotypic constraints of the cell.
        To obtain this approximation function, information regarding phenotypic constraints, as represented by PCs, is required in advance.
        However, in a real cell, it may not be easy to determine this information: a large number of PCs are required to provide phenotypic constraints.
        Here, we propose a simpler alternative measure for predicting cross-fitness and demonstrate its reliability using $\tilde{P}=3$ conditions.
    
        Instead of the difference between the PCs of phenotypic changes in response to environmental stresses, we adopted a simple measure between two phenotypic responses to environmental stresses: cosine-similarity for phenotypic change $\boldsymbol{\delta x^*(E)}$ in response to the stress environment $\boldsymbol{E^{(1)}}, \boldsymbol{E^{(2)}}$ defined as follows:
        
        \begin{equation}
            S_{c}(\boldsymbol{E^{(1)}}, \boldsymbol{E^{(2)}}) = \frac{\left(\boldsymbol{\delta x^*(E^{(1)})}\cdot\boldsymbol{\delta x^*(E^{(2)})}\right)}{\|\boldsymbol{\delta x^*(E^{(1)})}\|\|\boldsymbol{\delta x^*(E^{(2)})}\|}.
        \end{equation}
    
        This is a quantity characterizing orientations between phenotypic changes $\boldsymbol{\delta x^*(E^{(1)})}$ and $\boldsymbol{\delta x^*(E^{(2)})}$; it takes 1 if they are oriented in the same direction, -1 if they are oriented in the exact opposite direction, and 0 if they are uncorrelated\footnote{Here, we used cosine-similarity as a measure of similarity of phenotypic change. A similar result can be obtained by using the correlation coefficient instead. Considering the correspondence with the results in Chapter 2, we consistently use the cosine-similarity in the present paper.}.
        The cosine-similarity is symmetric for the stress environments $\boldsymbol{E^{(1)}}$ and $\boldsymbol{E^{(2)}}$.
    
\begin{figure*}[hbt]
            \centering
            \includegraphics[width=\hsize]{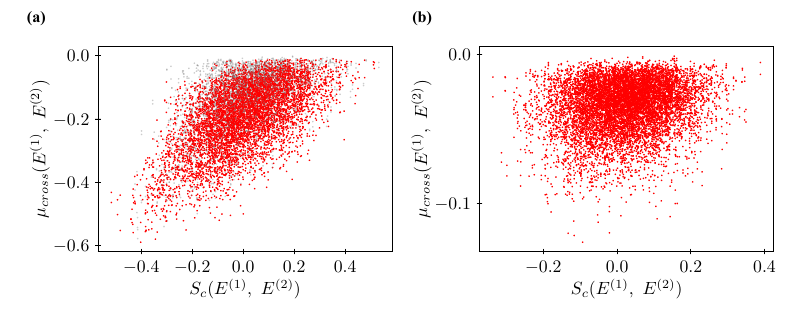}
            \caption{Cross-fitness is plotted against cosine-similarity.
            (a) $\tilde{P}=3$ case with 3-dimensional phenotypic constraint.
            Red dots correspond to the data between pairs of stress environments, under which the magnitude of fitness change was larger than 0.1, whereas gray dots include those smaller than 0.1.
            The correlation coefficient across all data is 0.57, and that across red dots only is 0.79.
            (b) $\tilde{P}=0$ case with no phenotypic constraint. The correlation coefficient is 0.25.
            }
            \label{fig:cos_vs_rc}
        \end{figure*}

        In Fig.\ref{fig:cos_vs_rc}(a), the cross-fitness is plotted against cosine-similarity across the pairs of randomly generated environments (both red and gray points).
        For $\tilde{P}=3$, one can see the correlation between cross-fitness and cosine-similarity (correlation coefficient 0.57).
        However, the correlation might not be significant, as shown in Fig.\ref{fig:cos_vs_rc}(a).
        The main reason for this is that for some stresses $\boldsymbol{E}$, the response is rather small, so the cosine-similarity and fitness change are small.
        To eliminate such "non-response" cases, we replotted the data across only the environment pairs under which the fitness decrease was larger than 0.1 (red dots in Fig.\ref{fig:cos_vs_rc}(a)), for which the correlation coefficient was 0.79.
    
        The prediction of cross-fitness using cosine-similarity does not require direct information on phenotypic constraints.
        However, such constraints are necessary for the correlation between cross-fitness and cosine-similarity.
        Owing to the low-dimensional constraint, the environmental and evolutionary responses are correlated in the low-dimensional space, which reflects the cosine-similarity (see also the discussion in the next section).
        For $\tilde{P}=0$, in which no phenotypic constraints evolved as no input-output relationship was postulated, such a correlation was not observed (Fig.\ref{fig:cos_vs_rc}(b), correlation coefficient 0.25).
    
\section{Representation of cross-fitness by fitness potential function}
\label{sec:potential_approximation_to_cross_fitness}

    \subsection{Potential approximation of cross-fitness in low-dimensional phenotype space}
    \label{ssec:Potential_approximation_of_cross-fitness_in_low-dimensional_phenotype space}

        In the previous section, we showed that cross-fitness can be approximated based on the information that the phenotypic response of cells to environmental stresses is constrained in low-dimensional space.
        In this section, we describe a potential theory that characterizes the phenotypic response by representing fitness as a function of environmental and genetic changes.
        
        For this, we consider a fitness function $u(\boldsymbol{X})$ of $D$-dimensional variable $\boldsymbol{X}=(X_1,X_2,\dots,X_D)$.
        $\boldsymbol{X}$ is given as a function of genotype $\boldsymbol{G}$ and environment $\boldsymbol{E}$.
        When phenotypic constraints exist, the phenotypic changes caused by environmental stress and genotypic mutations are restricted to a low, $D$-dimensional submanifold within the total $N$-dimensional phenotypic space.
        
        We assume that the fitness function $u(\boldsymbol{X(G, E)})$ has a maximum value at $\boldsymbol{G^{(0)}},\boldsymbol{E^{(0)}}$.
        This means that the cell with genotype $\boldsymbol{G^{(0)}}$ is adapted to environment $\boldsymbol{E^{(0)}}$.
        By expanding the fitness function $u(\boldsymbol{X(G, E)})$ around $\boldsymbol{G=G^{(0)}, E=E^{(0)}}$ up to the second order, we obtain the following:
        \begin{subequations}
            \begin{align}
                u &\simeq u_0 + \delta u^{(2)}, \label{eq:2dim_growth_rate} \\*
                u_0 &= u(\boldsymbol{X(G^{(0)}, E^{(0)}})) \\*
                \delta u^{(2)}(\boldsymbol{\delta X}) &= \frac{1}{2}\sum_{i, j}\frac{\partial^2 u}{\partial X_i\partial X_j}\delta X_i\delta X_j \label{eq:2dim_growth_rate_2}\\*
                \delta X_i(\boldsymbol{\delta G, \delta E} ) &= \delta X_i^{\boldsymbol{G}}(\boldsymbol{\delta G})+\delta X_i^{\boldsymbol{E}}(\boldsymbol{\delta E}),
            \end{align}            
        \end{subequations}
        where $\delta X_i^{\boldsymbol{G}}(\boldsymbol{\delta G}) = \left[\partial X_i/\partial \boldsymbol{G}\right]^{\boldsymbol{G=G^{(0)}}}_{\boldsymbol{E=E^{(0)}}}\cdot\boldsymbol{\delta G}$ with $\boldsymbol{\delta G} = \boldsymbol{G} - \boldsymbol{G^{(0)}}$ and $\delta X_i^{\boldsymbol{E}}(\boldsymbol{\delta E}) = \left[\partial X_i/\partial \boldsymbol{E}\right]^{\boldsymbol{G=G^{(0)}}}_{\boldsymbol{E=E^{(0)}}}\cdot\boldsymbol{\delta E}$ with $\boldsymbol{\delta E} = \boldsymbol{E} - \boldsymbol{E^{(0)}}$.
        
        Note that the first-order derivatives $(\partial u/\partial \boldsymbol{E})$ and $(\partial u/\partial \boldsymbol{G})$ are equal to $\boldsymbol{0}$, because the fitness function reaches a maximum at $\boldsymbol{G^{(0)}, E^{(0)}}$.
        Fitness decreases with environmental stress $\boldsymbol{E}$; however, it is recovered by changing genotype $\boldsymbol{G}$.
        At the end of the evolution, the fitness reaches a local maximum with genotype $\boldsymbol{G^*(E)}$.
        Hence, $\boldsymbol{G^*(E)}$ should satisfy the following condition:
        
        \begin{equation}
            \boldsymbol{G^*(E)} = argmax_{\boldsymbol{G}}[u(\boldsymbol{X(G, E)})].
        \end{equation}
        
        By expanding the above equation to the second-order terms of $\boldsymbol{\delta G}$ and $\boldsymbol{\delta E}$, we obtain:
        
        \begin{equation}
            \boldsymbol{G^*(E)} \simeq argmax_{\boldsymbol{G}}\left[\sum_{i, j}\frac{\partial^2 u}{\partial X_i\partial X_j}\delta X_i\delta X_j\right],
            \label{eq:2dim_maximum}
        \end{equation}
        where $\delta X_i = \delta X_i(\boldsymbol{G^*(E), E})$ and $\boldsymbol{\delta G^*(E) = G^*(E) - G^{(0)}}$.
        
        Because the fitness function takes the maximum value at $\boldsymbol{G^{(0)}, E^{(0)}}$, all eigenvalues of the matrix $\boldsymbol{H} = \{H_{ij}=(\partial^2 u/\partial X_i\partial X_j)_{\boldsymbol{G^{(0)}, E^{(0)}}}\}$ are negative, and $\boldsymbol{H}$ satisfies $\boldsymbol{x^T H x} \leqq 0$ for any vector $\boldsymbol{x}$, $\boldsymbol{x^THx}$ takes 0 and only under the condition $\boldsymbol{x = 0}$ (we assume that the change by stress is not so large, and remains within the range of the linear approximation in Eq.(\ref{eq:2dim_growth_rate}) is valid). 
        Therefore, when fitness is completely recovered by the genotype change $\boldsymbol{G^{(0)}\rightarrow G^*(E)}$, $\boldsymbol{\delta X(\delta G^*(E), \delta E)}$ should be a zero vector.
        Then, the following relationship holds:
        
        \begin{equation}
            \delta X_i^{\boldsymbol{G}}(\boldsymbol{\delta G^*(E)}) = -\delta X_i^{\boldsymbol{E}}(\boldsymbol{\delta E}) \ (i=1,2,\dots,D).
            \label{eq:cancelout_phenotypicchange}   
        \end{equation}
        
        Here, $(\partial \boldsymbol{X}/\partial \boldsymbol{E})_{\boldsymbol{G^{(0)}, E^{(0)}}}\cdot \boldsymbol{\delta E}$ is the first-order approximation of the phenotypic changes when the cell with genotype $\boldsymbol{G^{(0)}}$ is subjected to environmental change $\boldsymbol{E^{(0)}}\rightarrow\boldsymbol{E}$.
        The conditions Eq.(\ref{eq:cancelout_phenotypicchange}) indicate that the phenotypic changes caused by environmental change $\boldsymbol{E^{(0)} \rightarrow E}$ are cancelled out by genetic changes $\boldsymbol{G^{(0)}\rightarrow G^*(E)}$ (justification of Eq.(\ref{eq:cancelout_phenotypicchange}) is discussed in Sec.S6).
        Note that the existence of a genotype $\boldsymbol{G^*(E)}$ may not always be guaranteed.
        However, as the dimension of the genotypic space is much larger than that of the phenotypic space, such a genotype will exist.
        
        When the conditions Eq.(\ref{eq:cancelout_phenotypicchange}) are satisfied, then Eq.(\ref{eq:2dim_growth_rate_2}) with $\boldsymbol{G^*(E^{(1)})}$ and $\boldsymbol{E^{(2)}}$ can be written as follows:
            
        \begin{subequations}
            \begin{align}
                \delta u^{(2)} &= \frac{1}{2}\sum_{i, j}\frac{\partial^2 u}{\partial X_i\partial X_j}\delta X_i'\delta X_j', \label{eq:CrossFitness_2dim} \\
                \delta X_i'(\boldsymbol{\delta E^{(1)}, \delta E^{(2)}}) &= \delta X_i^{\boldsymbol{E}}(\boldsymbol{\delta E^{(1)}}) - \delta X_i^{\boldsymbol{E}}(\boldsymbol{\delta E^{(2)}}).
            \end{align}
        \end{subequations}
        
        This equation implies that the fitness of the cell that has evolved under environment $\boldsymbol{E^{(2)}}$, placed in environment $\boldsymbol{E^{(1)}}$, is given as a function of the difference between the phenotypic changes $\boldsymbol{\delta X^E(\delta E)}$ under environments $\boldsymbol{\delta E^{(1)}}$ and $\boldsymbol{\delta E^{(2)}}$.
        In Sec.\ref{ssec:cross_fitness}, the change $\boldsymbol{\delta X^E}$ in Eq.(\ref{eq:CrossFitness_2dim}) is given by the changes in the PCs.
        Then, cross-fitness  $\mu_{cross}\boldsymbol{(E^{(1)}, E^{(2)})} = \mu_0(\boldsymbol{E^{(2)}})\ \text{with} \ \boldsymbol{G=G^*(E^{(1)})})$ can be approximated as a function of the difference in the PC changes in the phenotypes between $\boldsymbol{E^{(1)}}$ and $\boldsymbol{E^{(2)}}$.

    \subsection{Application of potential theory to the result of evolution simulation}
    \label{ssec:Application_of_potential_approximation_to_the_result_of_evolutional_simulation}

        Following the argument in the last section, we estimate the coefficient of the second term of the cross-fitness $(\partial^2 \mu_{cross}/\partial y_1^2)$ ($y_1$ is the first PC of the phenotypic change $\boldsymbol{\delta x^*}$) from the change in $\mu_{cross}$ against the change in $y_1$.
        The solid lines in Fig.\ref{fig:Kernel} are predicted curves according to the above theory for $\tilde{P}=1$ and $\alpha=0.45$.
        The coefficient of the blue one is calculated by the least-squares method with $\mu_{cross} = -c y_1^2$ changing $c$.
        This curve approximates cross-fitness well, especially in regions where phenotypic changes are not too large.
        
        Next, We interpret the relationship between cosine-similarity and cross-fitness in Fig.\ref{fig:cos_vs_rc}(a) with the potential theory.
        By expanding $\delta X_i'$, Eq.(\ref{eq:CrossFitness_2dim}) can be rewritten as 
        
        \begin{equation}
            \begin{split}
                \delta u^{(2)} &=  \frac{1}{2}\sum_{i,j}\frac{\partial^2 u}{\partial X_i\partial X_j}\delta X^E_i(\boldsymbol{\delta E^{(1)}})\delta X^E_j(\boldsymbol{\delta E^{(1)}})\\
                &+ \frac{1}{2}\sum_{i,j}\frac{\partial^2 u}{\partial X_i\partial X_j}\delta X^E_i(\boldsymbol{\delta E^{(2)}})\delta X^E_j(\boldsymbol{\delta E^{(2)}})\\
                &- \sum_{i,j}\frac{\partial^2 u}{\partial X_i\partial X_j}\delta X_i^E(\boldsymbol{\delta E^{(1)}})\delta X_j^E(\boldsymbol{\delta E^{(2)}}).
            \end{split}
            \label{eq:Cross_fitness_indiv}
        \end{equation}
        
        The first and second terms in the above equation can then be interpreted as second-order approximations of fitness changes under stress environments $\boldsymbol{E^{(1)}}$ and $\boldsymbol{E^{(2)}}$.
        The third term represents the interaction between stress environments $\boldsymbol{E^{(1)}}$ and $\boldsymbol{E^{(2)}}$, that is, the fitness change in $\boldsymbol{E^{(2)}}$ owing to adaptive evolution under $\boldsymbol{E^{(1)}}$.
        This term is proportional to the inner product of the phenotypic changes $\boldsymbol{\delta X^E(\delta E^{(1)})}$ and $\boldsymbol{\delta X^E(\delta E^{(2)})}$ under the metric $\boldsymbol{H}=\{\partial^2 u/\partial X_i \partial X_j |_{\boldsymbol{G^{(0)}, E^{(0)}}}\}$.
        In other words, the third term corresponds to the difference in the orientation of phenotypic changes in $\boldsymbol{E^{(1)}}$ and $\boldsymbol{E^{(2)}}$; the similarity between the environments.
        In particular, when the first and second terms take the same value (fitness changes in $\boldsymbol{E^{(1)}}$ and $\boldsymbol{E^{(2)}}$ are the same), the cross-fitness is proportional to the cosine-similarity under the metric $\boldsymbol{H}$.
        This proportional relationship between cross-fitness and cosine-similarity supports the results presented in Sec.\ref{ssec:prediction_cross-resistance}.
        Note that this relationship is obtained because the Hessian matrix of the cross-fitness can be approximated well by a constant multiple of the unit matrix in the present model.
    
\section{Application for laboratory evolution of resistance to antibiotics}
\label{sec:application_for_experimental_data}
    
    In this section, we apply the present theory to the experimental evolution of antibiotic resistance in \textit{E.coli}, to confirm the prediction of cross-fitness by transcriptome changes.
    For details of the experiment evolution, see Sec.\ref{sec:materials_and_method} and \cite{suzuki2014prediction}.

    Here, 6 antibiotics $A^{(1)}, A^{(2)},\dots, A^{(6)}$ were used for the experimental evolution.
    First, each antibiotic $A_k$ was added to the parental strain up to the level as long as the growth is sustained.
    Indeed this level is called the Minimum Inhibitory Concentration (MIC), which is the lowest concentration of an antibiotic that prevents visible bacterial growth, and is used as a measure to quantify antibiotic resistance, which corresponds to the fitness here.
    As a measure of phenotypic changes, we used log-transformed transcriptome responses, following our previous study \cite{kaneko2015universal}, because changes in gene expression typically occur on the logarithmic scale.
    Namely, the phenotypic change was measured by the transcriptome change as $\delta X_i(A^{(k)})=\log_2(x_i(A^{(k)})/x_i(ND))$, where $\boldsymbol{x}(A_k)$ is the transcriptome data when an antibiotic $A_k$ is added near the MIC to the parent strain before evolution and $\boldsymbol{x}(ND)$ is the geometric mean of 3 independently measured transcriptome data under no-drug condition. 
    As the fitness measure, we used log-transformed MIC values [$\log 2(\mu$ g/ml] based on the previous study \cite{suzuki2014prediction}, which showed a linear correlation between log-transformed transcriptome changes and log-transformed MIC values.
    As MIC is larger, the fitness under the antibiotic is larger, so the former can be used as a measure of fitness
    Assuming that the laboratory evolution results in complete adaptation to antibiotics, we computed the relative MIC $R_{MIC}(A^{(k)}, A^{(l)})=MIC(A^{(k)}, A^{(l)}) - MIC(A^{(l)}, A^{(l)})$, where $MIC(A^{(k)}, A^{(l)})$ is the log-transformed MIC for $A^{(l)}$ of the strain that evolved to be resistant to $A^{(k)}$, and used it as the measure of cross-fitness \footnote{The relative MIC $R_{MIC}(A^{(k)}, A^{(l)})$ is the quantity corresponding to the cross-fitness in this paper. Cross-resistance $r_{MIC}(A^{(k)}, A^{(l)})$ is represented as $r_{MIC}(A^{(k)}, A^{(l)}) = R_{MIC}(A^{(k)}, A^{(l)}) - R_{MIC}(ND, A^{(l)})=MIC(A^{(k)}, A^{(l)})-MIC(ND, A^{(l)})$, which is consistent with commonly used cross-resistance.}.
    This quantity is non-positive which takes zero when $A^{(k)}$ and $A^{(l)}$ are the same antibiotics.
    
\begin{figure*}[hbt]
            \centering
            \includegraphics[width=\hsize]{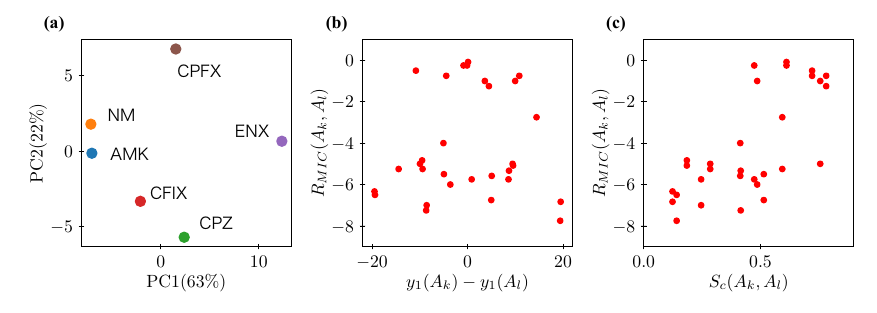}
            \caption{
            (a) Phenotypic changes $\boldsymbol{\delta X}(A^{(k)})\ (k=1,2,\dots,6)$ are plotted in PC plane.
            PCs are computed across the transcriptome changes against these antibiotics changes over 123 genes.
            (b) Relative MIC $R_{MIC}(A^{(k)}, A^{(l)})$ [$\log_2(\mu\text{g/ml})$] are plotted against the difference $y_1(A^{(k)}) - y_1(A^{(l)})$ of the first principal component of $\boldsymbol{\delta X}(A^{(k)})$.
            (c) Relative MIC $R_{MIC}(A^{(k)}, A^{(l)})$ [$\log_2(\mu\text{g/ml})$] are plotted against cosine-similarity $Sc(A^{(k)}, A^{(l)})$.
            $S_c(A^{(k)}, A^{(l)})$ is defined as $S_c(A^{(k)}, A^{(l)})=(\boldsymbol{\delta X}(A^{(k)})\cdot \boldsymbol{\delta X}(A^{(l)}))/|\boldsymbol{\delta X}(A^{(k)})||\boldsymbol{\delta X}(A^{(l)})|$.
            Pearson's correlation coefficient is 0.70 and the p-value is $1.5\times 10^{-5}$.
            }
            \label{fig:Suzuki_experiment}
    \end{figure*}

    We then performed principal component analysis (PCA) for the transcriptome changes $\boldsymbol{\delta X}(A^{(1)}), \boldsymbol{\delta X}(A^{(2)}), \cdots, \boldsymbol{\delta X}(A^{(6)})$ and plotted in PC plane (Fig.\ref{fig:Suzuki_experiment}(a)).
    The contribution of the first principal component accounts for a high percentage of $63\%$.
    Then, similarly to Sec.\ref{ssec:prediction_cross-resistance}, we computed the differences $y_1(A^{(k)})-y_1(A^{(l)})$ of the first principal component for an antibiotic $A^{(k)}$ used for the evolution of resistance and antibiotics $A^{(l)}$ used to measure the resistance of the evolved strain.
    In Fig.\ref{fig:Suzuki_experiment}(b), we plotted the measure of cross-fitness $R_{MIC}(A^{(k)}, A^{(l)})$ against $y_1(A^{(k)})-y_1(A^{(l)})$, which would correspond to Fig.\ref{fig:Kernel}(a).
    However, a clear one-dimensional curve as in Fig.\ref{fig:Kernel}(a) was not observed.
    One possible explanation for this is that the phenotypic changes were not constrained in a 1-dimensional space.
    
    From these data, we plotted the correlation between $R_{MIC}(A^{(k)}, A^{(l)})$ and $S_c(A^{(k)}, A^{(l)})$ (Fig.\ref{fig:Suzuki_experiment}(c)), which corresponds to Fig.\ref{fig:cos_vs_rc}(a), showing a significant correlation between them (Pearson's correlation coefficient is 0.70 with a p-value of $1.5\times10^{-5}$).
    Recalling that the gene expression dynamics involve hundreds of genes, this value is remarkable.
    Indeed it is about a similarly high value as obtained from the simulation.
    This result indicates that the present theory applies to the laboratory evolution of resistance to antibiotics.

\section{Discussion}
\label{sec:discussion}

    In the present study, we first evolved the gene regulatory network to be capable of multiple input-output relationships, to demonstrate that phenotypic changes due to environmental stress and genetic mutations are constrained to a lower-dimensional subspace, whose dimension corresponds to the degrees of freedom of the input-output relationship required for fitness.
    Phenotypic changes due to environmental stress and genotypic mutation are constrained to the common subspace, as formulated by dynamical system theory (see also\cite{tlusty2017physical,xue2019environment,alba2021global,sakata2020dimensional,tang2020functional,husain2020physical,chuang2019homeorhesis} for the relevance of dimensional reduction in biological systems).
    
    In the present model, phenotypic constraints were caused by the separation of $\tilde{P}$ eigenvalues in the positive direction from the eigenvalues of the genotype $\boldsymbol{G}$.
    The inputs corresponding to these separated $\tilde{P}$ eigenvalues are amplified in the gene regulatory network, and the dominant phenotypic changes owing to environmental changes and genetic mutations are constrained in the directions of the eigenvectors of these eigenvalues.
    This suggests that the subspace of the phenotypic constraints reflects signals that have been important in the evolutionary process.
    
    Next, we conducted evolutionary simulations under stress environments $\boldsymbol{E^{(i)}}$ using the evolved GRN exhibiting phenotypic constraints as the ancestor cell.
    We defined the cross-fitness $\mu_{cross}(\boldsymbol{E^{(1)},E^{(2)}})$ as fitness of the cells evolved in the stress environment $\boldsymbol{E^{(2)}}$ in the stress environment $\boldsymbol{E^{(1)}}$.  
    We then demonstrated that this cross-fitness is represented by the $\tilde{P}$-dimensional PC of the phenotypic change $\boldsymbol{\delta x^*(E^{(1)})}$ and $\boldsymbol{\delta x^*(E^{(2)})}$, cellular state changes to each stress $\boldsymbol{E}$, before the evolution to stress.
    Thus, the cross-fitness $\mu_{cross}(\boldsymbol{E^{(1)}, E^{(2)}})$ can be well represented by low-dimensional ($\tilde{P}$) phenotypic variables.
    This indicates that the fitness of the stress environment can be predicted by measuring the phenotypic changes in pre-evolutionary cells by the application of environmental stress.
    
    To provide an approximate estimation of the cross-fitness observed in evolutionary simulations, we introduced the fitness potential.
    The decrease in fitness due to stress is recovered when the phenotypic change in evolution completely cancels out the phenotypic change due to the stressful environment.
    By using the potential theory, cross-fitness can be approximated by the difference in phenotypic responses to the stress used for the evolution and that applied later for the test.
    
    In this potential approximation, the cross-fitness is symmetric against $\boldsymbol{E^{(1)}}$ and $\boldsymbol{E^{(2)}}$, that is, $\mu_{cross}(\boldsymbol{E^{(1)}},\boldsymbol{E^{(2)}}) = \mu_{cross}(\boldsymbol{E^{(2)}},\boldsymbol{E^{(1)}})$.
    Note that this is obtained by expanding the fitness up to the second order of phenotypic changes due to environmental changes and genetic mutations by assuming that these are not very large.
    If perturbations are much larger, the third- or higher-order effects are not negligible, and the above symmetry no longer holds.
    However, even if the cross-fitness is not symmetric, it is expected to correlate with the cosine-similarity (which is symmetric by definition) to a sufficient degree.
    
    Notably, in the model in the present study, the Euclidian distance between the expression pattern of the output gene and the target pattern is used as the fitness function, which is symmetric for the input-output relationship $(\boldsymbol{\eta^{(0)}, t^{(0)}})$.
    This symmetry eliminates the third-(or odd-) order terms in the potential form.
    Hence, the symmetry could be violated to some degree, depending on the choice of the input-output relationship and fitness function.
    
    Finally, we discuss the applications of our theory to experimental studies on laboratory evolution.
    In the present study, we first discuss the prediction of cross-fitness using the PCs of phenotypic changes in response to environmental change.
    To do this, information on the representation of the fitness by the PCs is needed in addition to information on phenotypic constraints, which may not be obtained directly from experimental data.
    Later, however, we demonstrated the correlation between cross-fitness and cosine-similarity in phenotypic changes in response to the stressful environment.
    In fact, the transcriptome changes of laboratory evolution of \textit{E. coli} confirm this correlation whereas fitness changes were not merely represented by the 1sr PC.
    The trend in fitness changes after evolution, thus, could be predicted by the transcriptome changes due to antibiotics before evolution(Sec.\ref{sec:application_for_experimental_data}).
    
    In the potential theory in the present study, we focused only on the full recovery of fitness via adaptive evolution in a stressful environment.
    However, in actual evolution, fitness may not be fully restored.
    We expect that our study will still provide relevant information in such cases, as long as phenotypic constraint exists and evolution occurs along it under a given fitness landscape.
    However, when a single mutation introduces drastic phenotypic changes or strong epistasis occurs during evolution, the correlation between cross-fitness and transcriptome cosine-similarity may not be clearly observed.
    
    In the present study, long-term evolution provided a phenotypic constraint to the cellular state, and later adaptive evolution of such cells to a stress environment follows the already created constraint.
    This implicitly assumes that there is a time-scale gap between the evolution of the present cells and their laboratory evolution to gain stress tolerance.
    The phenotypic constraint itself was shaped by the former evolutionary process but was not altered by the latter, shorter-term evolutionary process.
    
\begin{figure}[hbt]
        \centering
        \includegraphics[width=\hsize]{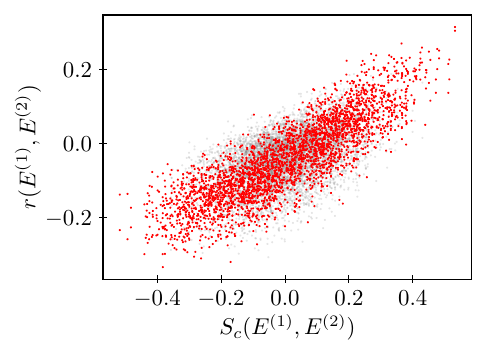}
        \caption{Cross-resistance is plotted against cosine-similarity for $\tilde{P}=3$ case with 3-dimensional phenotypic constraint.
        Red dots represent the data from the stress environment, under which fitness change is larger than -0.1.
        Gray dots represent data with smaller fitness changes. The correlation coefficient across all data is 0.69, and that across only red dots is 0.85.}
        \label{fig:cross_resistance}
    \end{figure}

    Thus far, we have discussed cross-fitness in the presence of phenotypic constraints. 
    Another commonly adopted measure of relative changes by evolution and adaptation is cross-resistance $r(\boldsymbol{E^{(1)}},\boldsymbol{E^{(2)}})$, given by
    
    \begin{equation}
        r(\boldsymbol{E^{(1)}}, \boldsymbol{E^{(2)}}) = \mu_{cross}(\boldsymbol{E^{(1)}}, \boldsymbol{E^{(2)}}) - \mu_{cross}(\boldsymbol{0}, \boldsymbol{E^{(2)}}).
        \label{eq:cross_resistance}
    \end{equation}
    It is defined as cross-fitness between $\boldsymbol{E^{(1)}}$ and $\boldsymbol{E^{(2)}}$ minus the fitness of the pre-evolutionary cell because we are mostly concerned with the relative fitness changes as a result of adaptive evolution.
    Therefore, even if the cross-fitness is symmetric, as in the present model, cross-resistance is not.
    We should note this point when applying the present theory to cross-resistance using Eq.(\ref{eq:cross_resistance}).
    Still, the present theory can be used to predict the cross-resistance (see Fig.\ref{fig:cross_resistance}) for an example of the correlation between cross-resistance and transcriptome cosine-similarity.
    This quantity is usually used as a measure to quantify antibiotic resistance.
    This will be useful for experimental verification of the present theory.
    
\section{Materials and Method}
\label{sec:materials_and_method}

    We describe the method of laboratory evolution and the acquisition of data used in Sec.\ref{sec:application_for_experimental_data}.
    The Insertion sequence-free {\it Escherichia coli} strain MDS42 \cite{posfai2017metabolic} was purchased from Scarab Genomics. 
    The cells were cultured in 200 $\mu$L modified M9 medium \cite{mori2011evaluating} in 96-well microplates with shaking at 900 strokes $\mathrm{min}^{-1}$ at 34${}^\circ$C.
    We prepared precultures by shaking -80${}^\circ$C glycerol stocked MDS42 strains for 23 h without antibiotics. 
    The cells precultured were diluted to an OD$_{600~\mathrm{nm}}$ of 1 $\times$ 10$^{-4}$ into 200 $\mu$L of fresh modified M9 medium in 96-well microplates with and without antibiotics. 
    The final concentrations of antibiotics used in this study were as follows; 3.9 $\times 10^{-3}~\mu$g/mL for Cefoperazone (CPZ), 1.2 $\times 10^{-2}~\mu$g/mL for Cefixime (CFIX), 4.0 $\mu$g/mL for Amikacin, 2.0 $\mu$g/mL for Neomycin (NM), 3.1 $\times 10^{-2}~\mu$g/mL for Enoxacin (ENX), and 2.0 $\times 10^{-3}~\mu$g/mL for Ciprofloxacin (CPFX), respectively. 
    The cultures were grown to an OD$_{600~\mathrm{nm}}$ in the 0.072$\sim$0.135 range (the equivalent of 10 generations).
    180 $\mu$L of exponential cultures were withdrawn rapidly, and cells were killed immediately by the addition of an equal volume of ice-cold ethanol that contained 10\% (w/v) phenol. 
    The cells were collected by centrifugation at 20,000 $\times~g$ at 4${}^\circ$C for 5 min, and the pelleted cells were stored at -80${}^\circ$C prior to RNA extraction. 
    Total RNA was isolated and purified from cells using RNeasy micro Kit with on-column DNA digestion (Qiagen) in accordance with the manufacturer’s instructions.
    
    Transcriptome analysis was performed as a previous study \cite{suzuki2014prediction} by using a custom-designed Agilent 8 $\times$ 60 K array for {\it E. coli} W3110. 
    Briefly, 100 ng of each purified total RNA sample was labeled using the Low Input Quick Amp WT Labeling kit (Agilent Technologies) with Cyanine3 (Cy3) according to the manufacturer’s instructions. 
    Cy3-labeled cRNAs were fragmented and hybridized to the microarray for 17 h at 65${}^\circ$C in a hybridization oven.
    Washing and scanning of microarrays were performed in accordance with the manufacturer’s instructions
    Microarray image analysis was performed using Feature Extraction version 10.7.3.1 (Agilent Technologies).
    
    The MIC values of evolved \textit{E. coli} strains for the aforementioned 6 antibiotics were obtained in the previous study \cite{suzuki2014prediction}.
    The transcriptome data and MIC values are available upon request.

    In the data analysis, the intensity values were normalized using the quantile normalization method.
    We then excluded the following genes which the parent strain lacks \textit{fhuA, yagE, yagF, yagG, yagM,yagX, appY, ycdR, ymfD, ymfI, ycgG, paaJ, ydbD, cheW, yfjL, yqiG, yqiI, yhhZ, yrhA, intB, yjhI, fimD, hsdR}, and \textit{yjiY}.
    Furthermore, we excluded genes with low expression levels ($\leq 100$ a.u. in any strain) and with relatively small expression change in response to all 6 antibiotics ($\delta X_i(A_k)=\log_2(x_i(A_k)/x_i(ND))\leq 1$ for all $A_k$), since the expression changes of such low expression or relatively unchanged genes were dominated by the experimental errors.

\begin{acknowledgments}
    We thank Tetsuhiro Hatakeyama for his insightful comments. 
    This research was partially supported by a Grant-in-Aid for Scientific Research (A) 431 (20H00123), a Grant-in-Aid for Scientific Research on Innovative Areas (17H06386) from the Ministry of Education, Culture, Sports, Science and Technology (MEXT) of Japan, the Japan Society for the Promotion of Science (20J12168), and the Japanese Science and Technology agency (JST) ERATO (JPMJER1902). 
    This research was also supported by the Novo Nordisk Foundation.
\end{acknowledgments}

\appendix

\section{Cosine similarity between phenotypic changes due to environmental stress and genetic changes through adaptive evolution}
\label{app:cosine_similairity}
        
        In Sec.\ref{sec:potential_approximation_to_cross_fitness}, we assumed that phenotypic changes due to environmental changes would be completely canceled out by phenotypic changes due to genotypic changes through evolution.
        To test that this assumption held in the evolutionary simulations, we measured the cosine similarity between phenotypic changes against environmental stress and the evolution under it;
        
        \begin{equation}
            S_{c} (\boldsymbol{E})=\frac{(\boldsymbol{\delta x_{env}^*(E)}\cdot\boldsymbol{\delta x^*_{evo}(E)})}{\|\boldsymbol{\delta x_{env}^*(E)}\|\|\boldsymbol{\delta x^*_{evo}(E)}\|},
            \label{eq:similarity}
        \end{equation}
        
        where $\boldsymbol{\delta x^*_{evo}(E)}$ is the phenotypic change in response to environmental stress $\boldsymbol{E}$ and $\boldsymbol{\delta x^*_{evo}(E)}$ is the phenotypic change from evolution with environmental stress $\boldsymbol{E}$.
        
\begin{figure*}[bht]
            \centering
            \includegraphics[width=0.9\hsize]{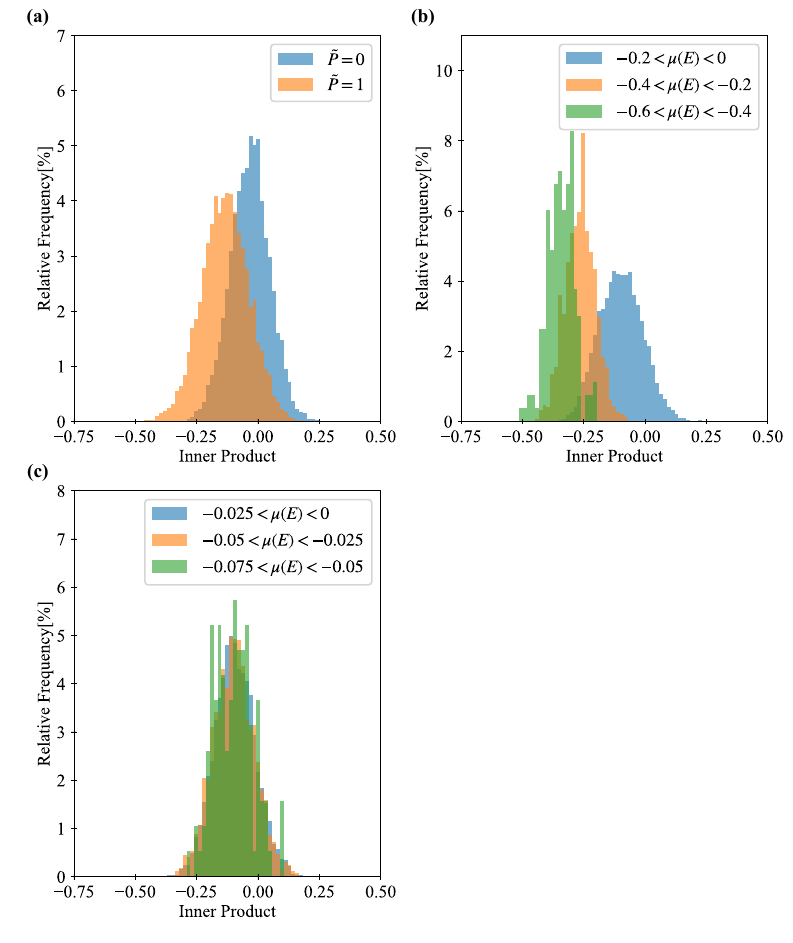}
            \caption{(a) Histogram of cosine similarity $S_{c}(\boldsymbol{E}) = (\boldsymbol{\delta x^*_{env}}(\boldsymbol{E}) \cdot \boldsymbol{\delta x^*_{evo}}(\boldsymbol{E}))/\|\boldsymbol{\delta x^*_{env}(E)}\|\|\boldsymbol{\delta x^*_{evo}(E)}\|$. The blue histogram represents the data of the genotypes evolved under $\tilde{P}=0$. The orange histogram represents the data for the genotypes that evolved under $\tilde{P}=1$, and we can see that the distribution shifts more negatively when there is a one-dimensional phenotypic constraint. (c) Histograms of (a) are divided into several parts according to the degree of fitness. (b) When $\tilde{P}=0$, the shape and position of the distribution do not change even if the fitness changes, but when $\tilde{P}=1$, the larger the change in fitness, the more the distribution shifts in the negative direction.}
            \label{fig:similarity}
        \end{figure*}

        The potential theory discussed in this section assumes the ideal limit in which the phenotypic change $\boldsymbol{\delta x_{evo}^*(E)}$ due to genotypic change will completely cancel out the phenotypic change $\boldsymbol{\delta x_{env}^*(E)}$ due to environmental change, thereby recovering the fitness.
        The histogram is shown in blue in Fig.\ref{fig:similarity}(a) represents the genotypes that evolved at $\tilde{P}=0, \alpha=0.45$.
        In this case, there are no phenotypic constraints, and the histogram does not deviate from the peak at similarity $ \sim 0$.
        In contrast, the histogram plotted in orange represents the data examined for adaptive evolution from genotypes that evolved at $\tilde{P}=1, \alpha=0.4$, where a one-dimensional phenotypic constraint is achieved.
        The distribution of cosine similarity is extended into negative regions, as shown in Fig.\ref{fig:similarity}(a).
        This implies that, when phenotypic constraints are present, the increased proportion of evolved cells has a cosine similarity closer to $-1$.
        The closer the cosine similarity is to $-1$, the more the phenotypic change $\boldsymbol{\delta x_{evo}^*(E)}$ in adaptive evolution is correlated with the phenotypic change $\boldsymbol{\delta x_{env}^*(E)}$ in response to environmental change.
        
        To investigate how a larger proportion of evolution leads to cosine similarity close to -1, the histograms in the presence of one-dimensional phenotypic constraints were computed separately according to fitness when the pre-evolutionary cells were subjected to environmental stresses.
        As plotted in Fig.\ref{fig:similarity}(b), the cosine similarity, as a result of evolution against environmental stresses, shifted to a negative value, heading towards -1 as the reduction in fitness inflicted on the pre-evolutionary cells was greater than that observed in the post-evolutionary cells.
        This tendency was not observed in the absence of phenotypic constraints (Fig.\ref{fig:similarity}(c)).
        
        In this model, the phenotypic space is 100-dimensional, whereas the number of output genes is eight. 
        Moreover, for genotypes evolved with $\tilde{P}=1$, the interaction matrix $\boldsymbol{O}$ between the gene regulatory network and the output gene is effectively a rank 1 matrix as a result of the evolution, with each row vector approximately 80\% correlated.
        This implies that a large number of phenotypic patterns capable of realizing a given target pattern exist.
        Accordingly, there is a huge variety of phenotypic changes that cancel out changes in fitness caused by environmental stress.
        When phenotypic constraints are present, phenotypic changes due to adaptive evolution are more likely to occur along these constraints in the direction opposite to that of the response to environmental stress.
        In fact, in Fig.\ref{fig:similarity}, the distribution of the cosine similarity extends to the negative region in the presence of phenotypic constraints.
        
\bibliography{Cross}% Produces the bibliography via BibTeX.

\end{document}

%%%%%%%%%% Merge with supplemental materials %%%%%%%%%%
\pagebreak
\widetext
\begin{center}
\textbf{\large Supplemental Materials: Prediction of Cross-Fitness for Adaptive Evolution to Different Environmental Conditions: Consequence of Phenotypic Dimensional Reduction}
\end{center}
%%%%%%%%%% Merge with supplemental materials %%%%%%%%%%
%%%%%%%%%% Prefix a "S" to all equations, figures, tables and reset the counter %%%%%%%%%%
\setcounter{equation}{0}
\setcounter{figure}{0}
\setcounter{table}{0}
\setcounter{page}{1}
\makeatletter
\renewcommand{\thesection}{S\arabic{section}}
\renewcommand{\theequation}{S\arabic{equation}}
\renewcommand{\thefigure}{S\arabic{figure}}
\renewcommand{\bibnumfmt}[1]{[S#1]}
\renewcommand{\citenumfont}[1]{S#1}
%%%%%%%%%% Prefix a "S" to all equations, figures, tables and reset the counter %%%%%%%%%%

\section{PCA on phenotypic changes caused by environmental stress $\boldsymbol{E}$}

\begin{figure*}[htb]
        \centering
        \includegraphics[width=\hsize]{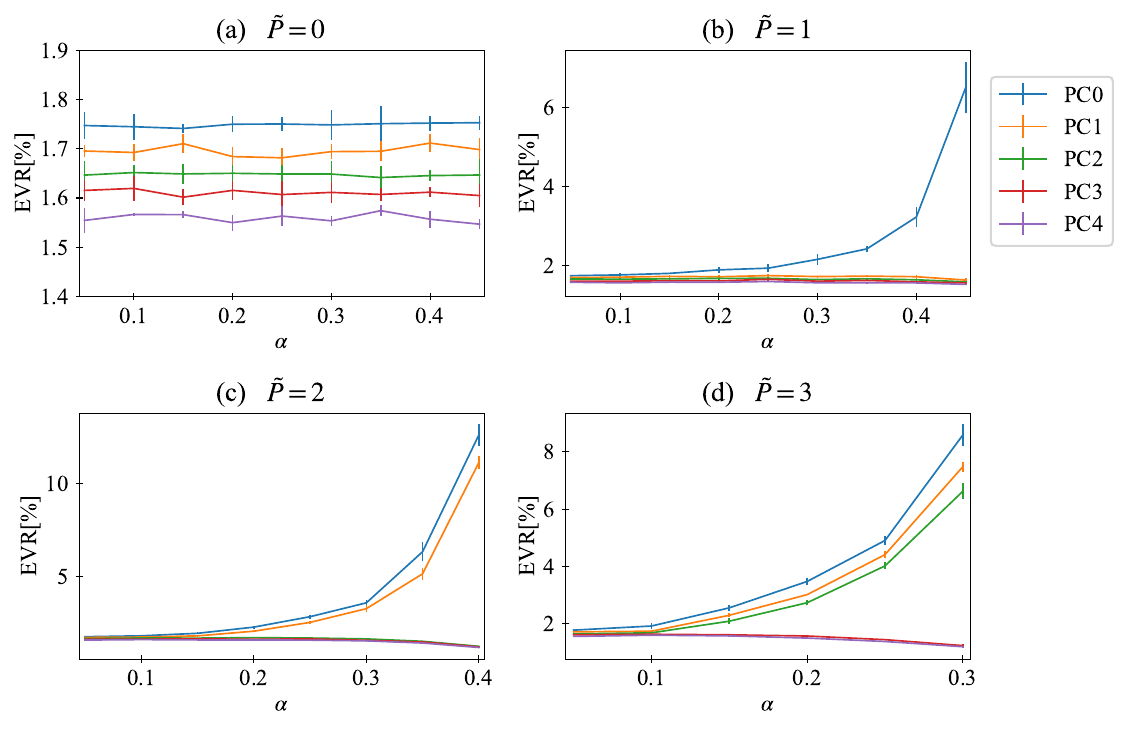}
        \caption{Explained variance ratio (EVR) of phenotypic changes for the first 5 principal components (PCs) when random environmental stresses $\boldsymbol{E}$ generated by $E_i\sim N(0,1)$ were applied.
        The changes in $\boldsymbol{x^*}$ for the evolved gene regulatory network were computed.
        The phenotypic changes $\boldsymbol{\delta x^*(E)}$ were obtained for 10,000 independent environmental stresses, from which PCs were computed.
        Each variance was calculated for the cell with the top fitness value in the population that evolved for (a) $\tilde{P} = 0$ (b) $\tilde{P} = 1$ (c) $\tilde{P} = 2$ (d) $\tilde{P} = 3$, and was plotted against the stress strength $\alpha$. The error bars represent the standard deviation of the five independent strains.}
        \label{fig:Var_Env}
\end{figure*}

\clearpage
\section{PCA on phenotypic changes caused by genetic mutations}

\begin{figure*}[htb]
    \centering
    \includegraphics[width=\hsize]{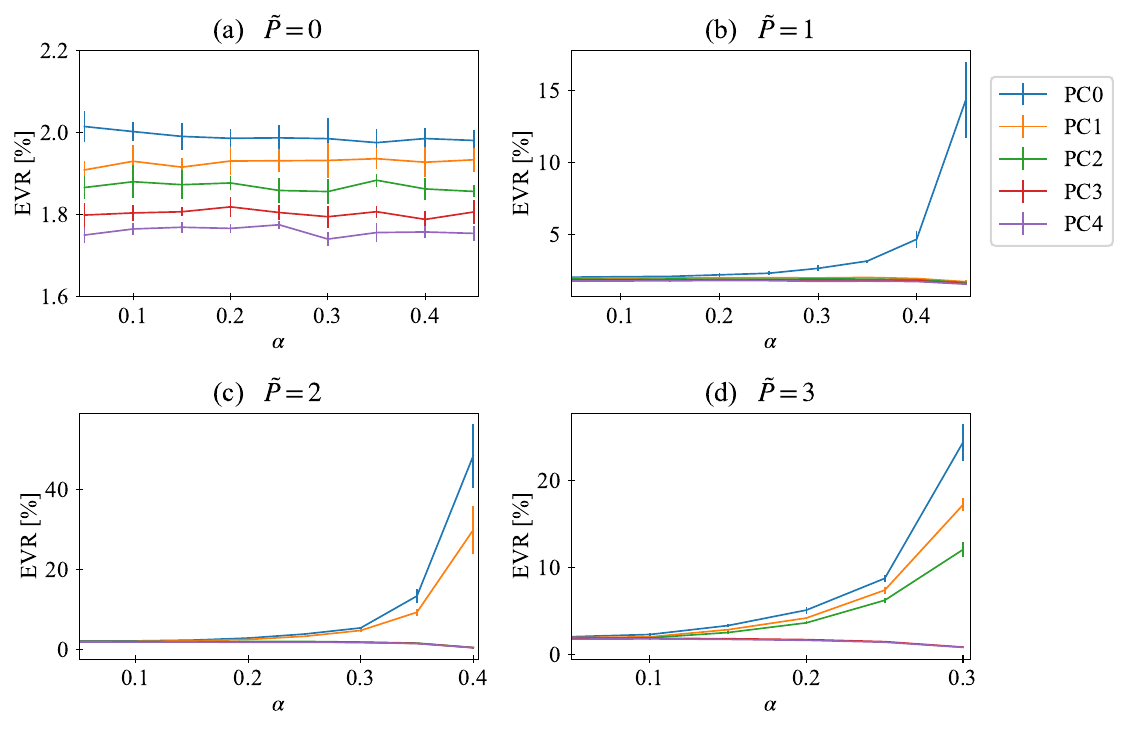}
    \caption{Explained variance ratio (EVR) of phenotypic changes due to genotypic changes.
    The genotypic change $\boldsymbol{G_0\rightarrow G_0+\delta G}$ was generated in the same manner in the evolution simulation with mutation rate $\rho=10/N^2=0.001$.
    10,000 phenotypic changes $\boldsymbol{\delta x^*(\delta G)}$ were used in the principal component analysis (PCA).
    Each variance was calculated for the cell with the top fitness value in the population that evolved at (a) $\tilde{P} = 0$ (b) $\tilde{P} = 1$ (c) $\tilde{P} = 2$ (d) $\tilde{P} = 3$. The error bars represent the standard deviation of the five independent strains.}
    \label{fig:Var_geno}
\end{figure*}

\clearpage

\section{Phenotypic Constraint of the present model in dynamical system}
\label{sec:phenotypic_constraint_in_dynamical_system}

    The formulation of phenotypic constraints in dynamical systems is given in \cite{sato2020evolutionary}, extending those in \cite{kaneko2015universal,furusawa2018formation}.
    In this chapter, we give the specific formula for the gene regulatory network in this paper.
    First, we restate the rate equations the gene regulatory network model in this paper follows.
    
    \begin{subequations}
        \begin{align}
            \dot{x_i} &= f_i(\boldsymbol{x}) - x_i \\
            f_i(\boldsymbol{x}) &= sig(y_i)=\frac{1}{1+\exp(-y_i)} \label{eq:sigmoid_function_app}\\
            y_i &= \frac{1}{\sqrt{N}}\sum_jG_{ij}x_j+\frac{1}{N_I}\sum_jI_{ij}\eta_j+E_i. \label{eq:interaction_GRN_app}
        \end{align}
        \label{eq:rate_equation_app}
    \end{subequations}
    
    The phenotype $\boldsymbol{x^*}$ of the cell is given as a fixed point in Eq.(\ref{eq:rate_equation_app}).
    Thus, the following equation is satisfied.
    
    \begin{equation}
        x_i^* = f_i(\boldsymbol{x^*}). \label{eq:fixed_point_app}
    \end{equation}
    
    Now, considering the total derivative of Eq.(\ref{eq:fixed_point_app}) by the parameter $s$
    
    \begin{equation}
        \frac{dx_i^*}{ds}=\frac{\partial sig(y_i^*)}{\partial s}+\sum_j\frac{\partial sig(y_i^*)^*}{\partial x_j^*}\frac{dx^*_j}{ds}.\label{eq:TotalDifference_1}
    \end{equation}
    
    Parameter $t$ corresponds to environmental stress $\boldsymbol{E}$ or genotype $\boldsymbol{G}$ in this paper.
    By rearranging Eq.(\ref{eq:TotalDifference_1}), we obtain the following equation.
    
    \begin{equation}
        \sum_j\left[\delta_{ij} - \frac{\partial sig(y^*_i)}{\partial x_j^*}\right]\frac{d x_j^*}{ds} = \frac{\partial sig(y_i^*)}{\partial s}.
        \label{eq:TotalDifference_2}
    \end{equation}
    
    Using the Jacobi matrix $\boldsymbol{J}$ and the susceptibility tensor $\boldsymbol{R_s}$ of the rate equation Eq.(\ref{eq:rate_equation_app}), Eq.(\ref{eq:TotalDifference_2}) can be rewritten in the following form.
    
    \begin{subequations}
        \begin{align}
            \frac{d x_i^*}{ds} &= -\sum_jJ^{-1}_{ij}[R_s]_j, \label{eq:TotalDifference_3}\\
            J_{ij} &= \left(\frac{\partial \dot{x_i}}{\partial x_j}\right)_{\boldsymbol{x=x^*}} = \frac{\partial sig(y_i^*)}{\partial x_j^*}- \delta_{ij}, \\
            \left[\boldsymbol{R_s}\right]_{i} &= \left(\frac{\partial \dot{x_i}} {\partial s}\right)_{\boldsymbol{x=x^*}} = \frac{\partial sig(y_i^*)}{\partial s}.
        \end{align}
    \end{subequations}
    
    Hence, within a linear approximation, the phenotypic change $\boldsymbol{x^*(\delta s)}$ due to the parameter change $s_0 \rightarrow s_0 + \delta s$ is
    
    \begin{equation}
        \boldsymbol{\delta x^*(\delta s)} \simeq \frac{d \boldsymbol{x^*}}{ds}\delta s=-\boldsymbol{J^{-1} R_s}\delta s.
        \label{eq:Relation_PhenotypicConstraint}
    \end{equation}
    
    Now consider the situation where the parameter change $\delta s$ satisfies $<\delta s>= 0, <\delta s\delta s'>= v\delta(\delta s, \delta s')$.
    The phenotypic changes $\boldsymbol{x^*}(\delta s)$ then satisfies
    
    \begin{subequations}
        \begin{align}
            <\boldsymbol{\delta x}(\delta s)> &\simeq \boldsymbol{0}\label{eq:average_phenotypic_change}\\
            <\boldsymbol{\delta x}(\delta s)[\boldsymbol{\delta x}(\delta s')]^T> &\simeq \boldsymbol{L R_s}<\delta s\delta s'> \boldsymbol{R_s^T L^T} \nonumber\\
            &= v\boldsymbol{L R_sR_s^T L^T}. \label{eq:variance_phenotypic_change}
        \end{align}
    \end{subequations}
    
    $\boldsymbol{L}=\boldsymbol{J^{-1}}$.
    Phenotypic constraint means that the phenotypic change $\boldsymbol{\delta x^*}(\delta s)$ of a cell due to a parameter change $s_0\rightarrow s_0+\delta s$ is restricted to a low-dimensional space.
    In other words, when an ensemble is taken with a random parameter change $\delta t$, the variances in the direction of certain eigenmodes become larger.

\section{Origin of dimensional reduction}
\label{sec:origin_of_dimension_reduction}

        In randomly generated genotypes in the present model, the response of the output genes to the environmental signal is quite small compared to the required response.
        Therefore, to achieve proper signal-target relationships, the input to the gene regulatory network must be amplified and transmitted to the target gene.
        Phenotypic constraints are acquired as a consequence of such evolution.
        This is suggested by the fact that the acquired phenotypes are constrained in the $\tilde{P}$-dimensional space depending on the magnitude of required output gene response $\alpha$.
        Under a linear approximation, the phenotypic changes $\boldsymbol{\delta x^*(E)}$ satisfies the following equations (see Sec.\ref{sec:phenotypic_constraint_in_dynamical_system} for derivation);
    
        \begin{align}
            <\boldsymbol{\delta x^*}(\boldsymbol{E})> &\simeq \boldsymbol{0}\label{eq:average_phenotypic_change_env}\\
            <\boldsymbol{\delta x^*}(\boldsymbol{E})[\boldsymbol{\delta x^*}(\boldsymbol{E'})]^T> &= \boldsymbol{L R_ER_E^T L^T}, \label{eq:variance_phenotypic_change_env}
        \end{align}
        where $\boldsymbol{L}=\boldsymbol{J^{-1}}$ is the inverse of the Jacobian matrix $\boldsymbol{J}$ in Eq.(\ref{eq:rate_equation_app}) around the fixed point and $\boldsymbol{R_E} = \{[\boldsymbol{R_E}]_{ij} = (\partial x_i/\partial E_j)\}$ is the partial derivative of the rate equation Eq.(\ref{eq:rate_equation_app}) against environmental stress $\boldsymbol{E}$.
        In the present model, $\boldsymbol{J}$ and $\boldsymbol{R_E}$ are as follows;
        \begin{subequations}
            \begin{align}
                J_{ij} &= (\partial \dot{x}_i/\partial x_j)|_{\boldsymbol{x^*}} = x_i^*(1-x_i^*)G_{ij} - \delta_{ij}\\
                \left[\boldsymbol{R_E}\right]_{ij} &= (\partial \dot{x_i}/\partial E_j)|_{\boldsymbol{x^*}} = x_i^*(1-x_i^*)\delta_{ij}
            \end{align}
        \end{subequations}
    
\begin{figure*}[hbt!]
            \centering
            \includegraphics[width=\hsize]{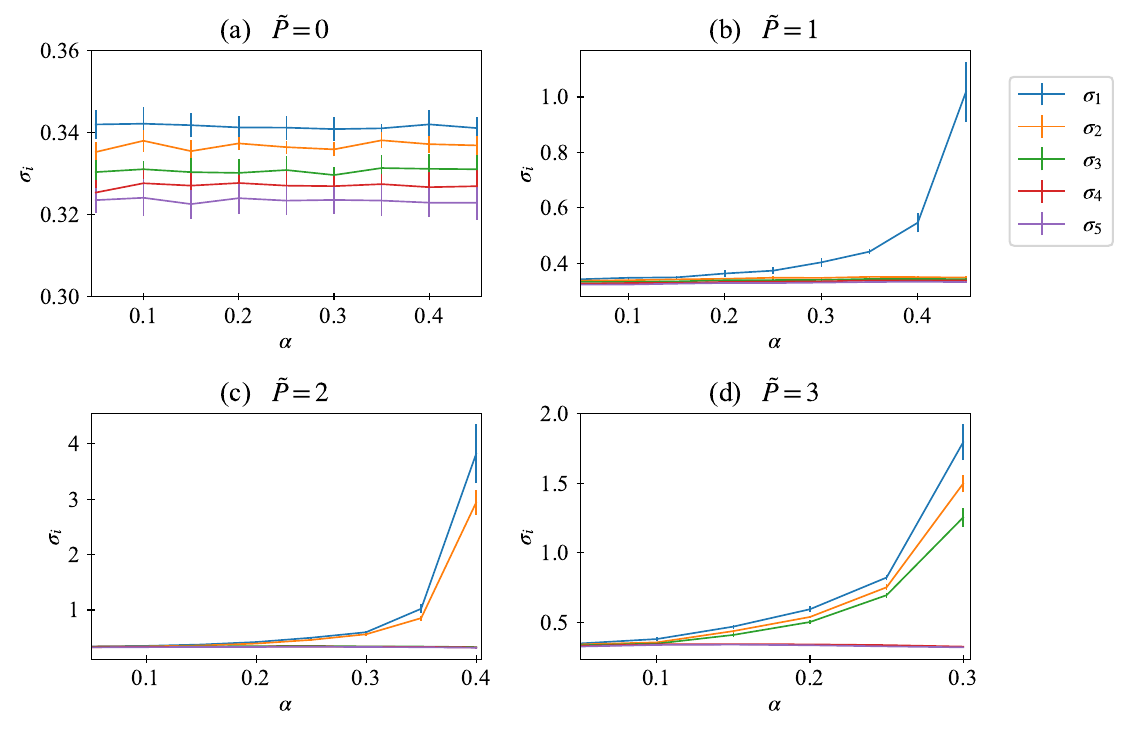}
            \caption{Singular values of $\boldsymbol{LR_E}$ around the phenotype $\boldsymbol{x^*}$ of the most highly adapted cells in the post-evolutionary population. Here, $\boldsymbol{L=J^{-1}}$ is the inverse of the Jacobi matrix and $\boldsymbol{R_E}=(\partial \dot{\boldsymbol{x}}/\partial \boldsymbol{E})$ is the response matrix to environmental stress $\boldsymbol{ E}$ is the response matrix to the environmental stress $\boldsymbol{E}$. Each singular value was calculated for the cell with the highest degree of adaptation in the population that evolved at (a)$\tilde{P}=0$\ (b)$\tilde{P}=1$\ (c)$\tilde{P}=2$\ (d)$\tilde{P}=3$.\ Here, the error bars represent the standard deviation in five independently evolved series.}
            \label{fig:Singular_value_LR}
        \end{figure*}

        This means that, within a linear approximation, the explained variance of the phenotypic change $\boldsymbol{\delta x^*(E)}$ with environmental change depends on the magnitude of the singular value of $\boldsymbol{L R_E}$.
        In fact, as shown in Fig.\ref{fig:Singular_value_LR}, the dependence of the singular value of $\boldsymbol{L R_E}$ on $\tilde{P}$ and $\alpha$ shows the same behavior as in Fig.\ref{fig:Var_Env}. This suggests that the phenotypic constraints observed in the simulations of this model can be explained by the linear mode around the fixed point of the dynamical systems (see Sec.\ref{sec:phenotypic_constraint_in_dynamical_system}).
    
\section{Eigenvalues of $\boldsymbol{G}$}
\label{sec:eigenvalue_of_J}
    
        Let us investigate the origin of the separation of $\tilde{P}$-dimensional components in terms of the eigenvalue structure of $\boldsymbol{G}$ that changes directly during evolution.
        In this model, the $\boldsymbol{G}$ of a pre-evolutionary cell was generated so that the diagonal elements are 0, and the off-diagonal components and chosen to be $-1$ or $1$ with equal probability. It is known that the eigenvalue distribution of such a random matrix is uniformly distributed within a circle of radius 1 centered at $N\rightarrow \infty$ (Circular Law)\cite{pastur2011eigenvalue}. 
            
\begin{figure*}[hbt!]
            \centering
            \includegraphics[width=\hsize]{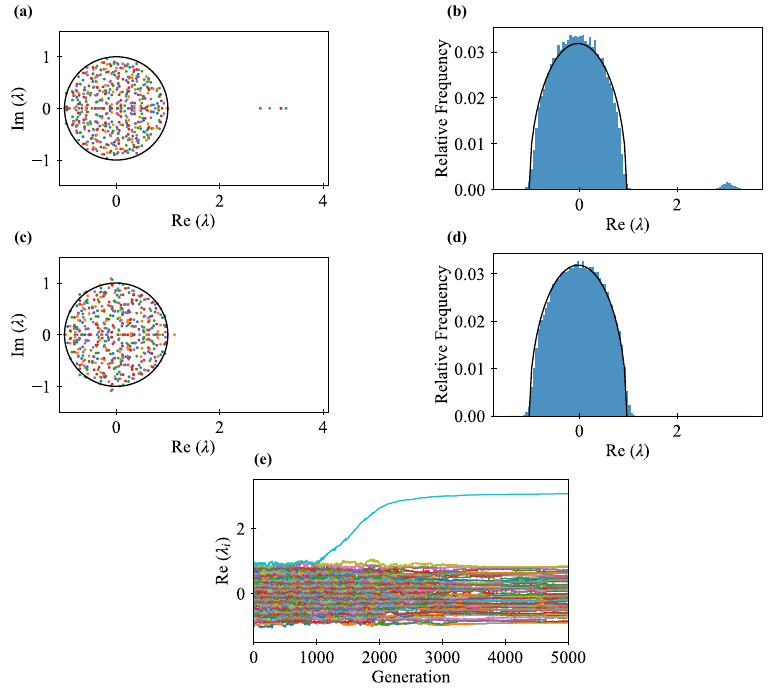}
            \caption{Eigenvalue distribution of genotype $\boldsymbol{G}$ at generation (a) 5000 and (c) 0 at $\tilde{P}=1$ and $\alpha=0.45$.
            The distribution calculated with independent evolutionary strains is plotted in different 5 colors.
            (b) and (d) are histograms of the relative frequency of the real part of eigenvalues.
            These were calculated across 500 cells (100 cells in 5 independent strains).
            (e) Evolutionary changes of the real parts of the eigenvalues of genotype $\boldsymbol{G}$ for the cell with the highest fitness. Plotted against the generation of evolution}
            \label{fig:genotype}
        \end{figure*}

        In the case of $N = 100$, used in this paper, the eigenvalue distribution of $\boldsymbol{G^{ini}}$ is almost uniformly distributed inside a circle of radius 1 centered at the origin.
        In the early stages of evolution, the phenotypic changes $\boldsymbol{x^*}$ under all input signal $\boldsymbol{\eta^{(m)}}$ are not sufficiently large enough that output genes make the corresponding target patterns (Fig.\ref{fig:genotype}(a)).
        On the other hand, the eigenvalue structure of the evolved genotypes on the complex plane is plotted in Fig.\ref{fig:genotype}(b).
        It can be seen that one eigenvalue is separated in the positive direction from the other eigenvalues. 
        Except for the first eigenvalue, the eigenvalues follow the uniform distribution within a circle of radius 1 centered at the origin, which is preserved as in $\boldsymbol{G^{ini}}$.
        The introduction of phenotypic constraints did not destroy the original structure of the random matrix in $\boldsymbol{G^{ini}}$ \cite{tao2013outliers}.
        This difference in the eigenvalue structures of the pre- and post-evolution cell was evolutionarily acquired.
        In Fig.\ref{fig:genotype}(c), the real part of the eigenvalues of $\boldsymbol{G}$ of the cell with the highest fitness is plotted across generations, for $\tilde{P} = 1$ and $\alpha = 0.45$.
    
        When $\tilde{P}=2$, two eigenvalues are separated from the circle, as in the case with $\tilde{P}=1$.
        Such preservation of the original eigenvalue structure is possible only when $\tilde{P}$ is small.
        As $\tilde{P}$ is larger, the radius of the circle in which the eigenvalues are distributed shrinks throughout evolution from the initial circle for a random matrix.
    
\bibliography{Cross}% Produces the bibliography via BibTeX.